\documentclass[11pt]{article}
\usepackage{jheppub}
\usepackage[dvipsnames]{xcolor}
\usepackage{simpler-wick}
\usepackage{xcolor}
\usepackage{amsmath}
\usepackage{amssymb}
\usepackage{bm}
\usepackage{amsfonts}
\usepackage{mathrsfs}
\usepackage{latexsym}
\usepackage{esint}
\usepackage{tikz}
\usepackage{tensor}
\usepackage{subcaption}
\usepackage{physics}
\usepackage{extarrows}
\usepackage{bbm}
\usepackage{graphicx} 
\usepackage{graphbox}
\usepackage{slashed}
\usepackage{mathtools}

\usepackage{orcidlink}

\usepackage{hyperref}
\hypersetup{
	colorlinks = True,
   	citecolor = NavyBlue,
   	linkcolor = NavyBlue,
	urlcolor = NavyBlue,
	bookmarks=true
}

\newcommand{\nn}{\nonumber}

\newcommand{\ldirac}{{\lbrack\!\!\!\;\langle}}
\newcommand{\rdirac}{{\rangle\!\!\!\;\rbrack}}

\DeclareMathOperator{\USp}{USp}
\DeclareMathOperator{\SO}{SO}

\newcommand{\ul}[1]{\underline{#1}}
\newcommand{\bul}[1]{\underline{\boldsymbol{#1}}}
\newcommand{\ra}{\rangle}
\newcommand{\la}{\langle}
\newcommand{\mbf}[1]{\mathbf{#1}}
\newcommand{\bs}[1]{\boldsymbol{#1}}

\newcommand{\vSpinor}{{\vphantom{\vert p_I\ra}}}

\title{Five-dimensional spinor helicity for all masses and spins}

\abstract{\!\!
We develop a spinor helicity formalism for five-dimensional scattering amplitudes of any mass and spin configuration.
While five-dimensional spinor helicity variables have been previously studied in the context of $\mathcal{N}=2,4$ supersymmetric Yang-Mills scattering amplitudes with spin less than two \cite{Chiodaroli:2022ssi}, we propose an alternative viewpoint that stems from $d$-dimensional spinor helicity variables avoiding the use of the exceptional low-dimensional isomorphism $\SO(4,1) \cong \USp(2,2)$ and the decomposition of a massive momentum into the sum of two massless momenta. 
By enumerating all possible independent little group tensors, we systematically build the full space of five-dimensional three-point tree-level scattering amplitudes for any configuration of spins and masses. 
Furthermore, we provide a prescription for computing the high energy limit of scattering amplitudes written in our spinor helicity variables.
We also expect that our formalism will be applicable to effective field theories with higher spin, in particular, the scattering of highly spinning black holes in 
five dimensions.
}

\author[a,\orcidlink{0000-0003-1186-4624}]{Andrzej Pokraka,}
\emailAdd{andrzej\_pokraka@brown.edu}

\author[a, \orcidlink{0000-0003-2827-2406}]{Smita Rajan,}
\emailAdd{smita\_rajan@brown.edu}

\author[a,\orcidlink{0000-0002-0846-8017}]{Lecheng Ren,}
\emailAdd{lecheng\_ren@brown.edu}

\author[a,b, \orcidlink{0009-0008-2506-3207}]{Anastasia Volovich,}
\emailAdd{anastasia\_volovich@brown.edu}

\author[c, \orcidlink{0000-0001-6192-7406}]{W. Wayne Zhao}
\emailAdd{wwzhao@princeton.edu}

\affiliation[a]{
    Department of Physics, 	
    Brown University, 	
    Providence, 	
    RI 02912, 
    USA
}

\affiliation[b]{Department of Physics,
	Harvard University,
 	Cambridge, MA 02138, USA}

  \affiliation[c]{Department of Physics,
	Princeton University,
 	Princeton, NJ 08544, USA}
\begin{document}
\maketitle

\section{Introduction}

It is well known that \emph{massless} on-shell scattering amplitudes are most naturally expressed in terms of spinor helicity variables illuminating many interesting structures that would otherwise be obscured by the usual presentation involving spacetime tensors and polarization vectors  \cite{Srednicki:2007qs, Schwartz:2014sze}. 
In particular, asymptotic massless states are uniquely specified by their helicity and momentum; spinor helicity variables make this manifest without introducing gauge redundancy. 
From this point of view, the spinor helicity framework represents amplitudes in their most physical form. 
In addition to conceptual advantages, the spinor helicity framework is an important computational tool that has facilitated the discovery of numerous important advances in the study of on-shell scattering amplitudes, starting with the celebrated Parke-Taylor formula for tree-level MHV amplitudes \cite{ParkeTaylor},
and all the further developments (see 
\cite{ElvangHuang, Badger:2023eqz, Travaglini:2022uwo, Bern:2022jnl}  
for reviews).
While the spinor helicity formalism revolutionized the study of massless scattering amplitudes four decades ago, spinor helicity formalisms for \emph{massive} particles remained largely underdeveloped.
It was only recently that a four-dimensional \emph{massive} any \emph{spin} spinor helicity formalism was developed by Arkani-Hamed, Huang and Huang \cite{Arkani-Hamed:2017jhn} (see also \cite{Craig:2011ws, Kiermaier:2011cr, Ochirov:2018uyq, Herderschee:2019dmc, Johansson:2019dnu, Chiodaroli:2021eug}).

By utilizing four-dimensional massive spinor helicity, Arkani-Hamed, Huang, and O’Con\-nell showed that graviton scattering with highly spinning Kerr black holes resembles the scattering of an elementary particle or, in other words, is minimally coupled \cite{Arkani-Hamed:2019ymq}. 
The authors also showed that the Janis-Newman shift
that relates Schwarzchild and Kerr black hole solutions has an interpretation as the exponentiation induced by taking the large spin limit of minimally coupled massive spinning particles (see \cite{Huang:2019cja, Chung:2019yfs, Emond:2020lwi} for generalizations to other four-dimensional solutions).
Since higher dimensional solutions to the Einstein equations can be obtained by similar generalized Janis-Newman shifts \cite{Erbin:2017}, it would be interesting to understand if these shifts also have an interpretation in terms of amplitudes.

Motivated by this question, we develop a five-dimensional spinor helicity formalism for any mass and spin. 
Our formalism is closely related to that of Chiodaroli,  Gunaydin, Johansson, and Roiban \cite{Chiodaroli:2022ssi} but differs in the sense that we do not expand the massive spinors in terms of massless spinors opting to directly work with spinors of the $\text{SO}(4)$ massive little group.
Moreover, we do not commit to a specific theory or Lagrangian and construct the space of independent three-point amplitudes from purely kinematic considerations. 
In particular, we find that the number of independent three-point amplitudes matches the expected counting from \cite{Costa:2011mg, Elkhidir:2014woa}. 
Lastly, we describe how to take the high energy limit of the massive amplitudes. 
We show that when the leading order term in the $m\ll E$ expansion vanishes, the subleading terms combine non-trivially to form a gauge invariant amplitude with $m=0$.

The paper is organized as follows.
In section \ref{sec:5Dformalism}, we describe our five-dimensional spinor helicity formalism and how it differs from \cite{Chiodaroli:2022ssi}. 
In section \ref{sec:3ptAmp}, we systematically construct all three-point amplitudes and verify that our construction produces the expected number of independent amplitudes.
In section \ref{sec:HElim}, we describe how to take the high energy limit of amplitudes built out of the spinor helicity variables. 


\section{Five-dimensional spinor helicity formalism\label{sec:5Dformalism}}

In this section, we construct a five-dimensional spinor helicity formalism for all masses and all spins. 
Five-dimensional massive spinor helicity formalism has been recently presented in \cite{Chiodaroli:2022ssi}.  They used
the isomorphism $\SO(4,1) \cong \USp(2,2)$ and decomposed massive momentum into the sum of two massless momenta.
We will not rely on these choices and use the formalism which we
hope could be generalized to $d\geq5$.
For the practical reader, this conceptual difference is essentially philosophical; it is easy to convert between our formalism and \cite{Chiodaroli:2022ssi}.

The Dirac spinor representation of the Lorentz group $\SO(4,1)$ is four-dimensional. 
We use gamma matrices from the $d$-dimensional definitions in appendix B of \cite{Polchinski:1998rr} with rows and columns 2 and 3 swapped.
This choice will make the form of the charge conjugation matrix simpler
\begin{align}
    &\Gamma^0= 
    \left(\begin{array}{cccc}
        0 & 0 & -1 & 0 \\
        0 & 0 & 0 & -1 \\
        1 & 0 & 0 & 0 \\
        0 & 1 & 0 & 0 
    \end{array}\right),
    \quad
    \Gamma^1 = 
    \left(\begin{array}{cccc}
        0 & 0 & -1 & 0 \\
        0 & 0 & 0 & 1 \\
        -1 & 0 & 0 & 0 \\
        0 & 1 & 0 & 0 
    \end{array}\right),
    \quad
    \Gamma^2 = 
    \left(\begin{array}{cccc}
        0 & 0 & 0 & 1 \\
        0 & 0 & 1 & 0 \\
        0 & 1 & 0 & 0 \\
        1 & 0 & 0 & 0 
    \end{array}\right)
    \nn\\&
    \Gamma^3 = 
    \left(\begin{array}{cccc}
        0 & 0 & 0 & -i \\
        0 & 0 & i & 0 \\
        0 & -i & 0 & 0 \\
        i & 0 & 0 & 0  
    \end{array}\right),
    \quad
    \Gamma^4 = 
    -i \Gamma^0 \Gamma^1 \Gamma^2 \Gamma^3 =
    \left(\begin{array}{cccc}
        1 & 0 & 0 & 0 \\
        0 & 1 & 0 & 0 \\
        0 & 0 & 1 & 0 \\
        0 & 0 & 0 & 1  
    \end{array}\right). 
\end{align}
These gamma matrices $(\Gamma^\mu)_A^{~~B}$ satisfy the usual Clifford algebra 
$\{\Gamma^\mu,\Gamma^\nu\}=2 \eta^{\mu \nu}$ with the mostly plus metric $\eta^{\mu\nu} = \mathrm{diag}(-1,1,1,1,1)$
and $\mu, \nu=0,1,2,3,4$.

The charge conjugation matrix $C$ obeying 
$ C \Gamma^\mu C^{-1} = + \Gamma^\top$ is given by 

\begin{equation}
 \label{cmatrix}
    C^{AB}
    = \left(\begin{array}{cccc}
        0 & -i & 0 & 0 \\
        i & 0 & 0 & 0 \\
        0 & 0 & 0 & i \\
        0 & 0 & -i & 0 
        \end{array}\right)
    \,.
   \end{equation}
We denote its inverse $C_{AB}$ by lowered indices  such that $C_{AB} C^{BC} = \delta_A^{\enspace C}$ and use it to lower and raise Dirac indices
$A,B=1,2,3,4$, as follows
\begin{equation}
    \lambda^A = \lambda_B C^{BA} ~~~~~
    \lambda_A = \lambda^B C_{BA}.
\end{equation}

The momentum space Dirac operator is 
\begin{align} \label{eq:pslash}
    p_A^{\enspace B} 
    = p_\mu (\Gamma^\mu)_A^{\enspace B}
    \,.
\end{align}
The on-shell condition is captured by the determinant $\det p_A^{\enspace B} \propto p^2 = -m^2$.
For massive particles $m\neq0$, $p_A^{\enspace B}$ is rank four while for massless particles $m=0$,  $p_A^{\enspace B}$ is rank two.

We will consider massless and massive cases separately in the next two subsections.

\subsection{Massless spinor helicity variables}

The Dirac operator \eqref{eq:pslash} can be factorized over the massless little group $\SO(3)$
\begin{equation} \label{eq:mless-Id}
    p_A^{\enspace B} 
    = \vSpinor_A \vert p_I \ra\
     \la p^I \vert^B
     \,,
\end{equation} 
where we sum over  the little group $\SO(3)$ indices $I=1,2$.

Massless spinor helicity (right-bracket) variables
$\vSpinor_A\vert p_I \ra$
satisfy massless Dirac equation
\begin{align}
     p_A^{\enspace B} \vSpinor_B\vert p_I \ra = 0.
\end{align}

We will use the following explicit representation \footnote{
Note that the mass dimension of these spinors is not uniform. 
While this choice may seem unnatural initially, it is extremely convenient since it preserves \eqref{eq:mless-Id} and avoids introducing unnecessary square roots into the parameterization of the spinors.
}
\begin{align} \label{eq:mless-spinors}
   \vSpinor_A\vert p_I \ra
   &=  
    \begin{pmatrix}
        (p_2 - i p_3) & -\frac{p_4}{p_0 + p_1} \\
        (p_0 + p_1) & 0 \\
        0 & 1 \\
        - p_4 & \frac{- p_2 - i p_3}{p_0 + p_1}
    \end{pmatrix}
    \,.
\end{align}

The corresponding left brackets are defined by contracting with the charge conjugation matrix \eqref{cmatrix}
\begin{align}
    \la p_I\vert^A = \vSpinor_{B} \vert p_I \ra\ C^{BA}
    .
\end{align}
Note that the position of the Dirac index determines whether the spinor helicity variable is written as a right or left bracket: lower Dirac indices for right brackets and upper Dirac indices for left brackets. 

The little group indices $I=1,2$ are raised and lowered with the
charge conjugation matrix 
\begin{equation}
 C^{IJ}=i \epsilon^{IJ}, ~~~ \epsilon^{12}=1 
\end{equation}
as follows
\begin{equation}
    \la p^I \vert^B=\la p_J \vert^B C^{JI}.
\end{equation}

With these conventions, spinor products are computed by contracting the spacetime-Dirac indices
\begin{align}
    {}\la p_I \vert q_J \ra
    = {}\la p_I \vert^A \vSpinor_A\vert q_J \ra
    = {}{\la p_I \vert}_A\ C^{AB}\ \vSpinor_B\vert q_J \ra.
\end{align} 
These spinor brackets are anti-symmetric by definition 
\begin{equation}
 {}\la p_I \vert q_J \ra = - {}\la q_J \vert p_I \ra.   
\end{equation}

Additionally, whenever there is a momentum $k$ sandwiched between two spinor brackets, it is interpreted as a $k_{A}^{\enspace B}$ acting on the brackets
\begin{align}
    {}\la p_I \vert k \vert q_J \ra
    = {}\la p_I \vert^A\ k_A^{\enspace B}\ \vSpinor_B\vert q_J \ra
    \,.
\end{align}
Moreover, to simplify formulas, we use the usual short-hand 
\begin{equation}
   \vSpinor_A\vert i_I \ra = \vSpinor_A\vert p_{iI} \ra 
\end{equation}
 to denote the spinor for the $i^\text{th}$ particle of an amplitude.

To simplify the process of keeping track of indices, we follow \cite{Chiodaroli:2022ssi} 
and contract the little group index of each particle with momentum $p_i$ with a dummy spinor to define the brackets
\begin{align} \label{eq:ul0}
    \vSpinor_A\vert \ul{i} \ra = \vSpinor_A\vert p_{iI} \ra \  z_i^I \,,
\end{align}

Spin-$S$ representations in five-dimensions correspond to little-group tensors that are fully symmetric in their $\SO(3)$ indices, and therefore contracting all little group indices into dummy spinors automatically imposes the correct symmetry and turns tensors into polynomials. 
Brackets are constructed from these underlined spinors in the same way as before by contracting the spacetime Dirac indices with the $C$-matrix
\begin{align}
    \la \ul{ik} \ra
    = {\la \ul{i} \vert}_A\  C^{AB}\ \vSpinor_B\vert \ul{j} \ra\,,
    \quad
    \la \ul{i}\vert k \vert \ul{l} \ra
    = {\la \ul{i} \vert}^A\ k_A^{\enspace B}\ \vSpinor_B\vert \ul{l} \ra
    \quad \text{ and so on.}
\end{align}
The actual tensor can be recovered by taking specific combinations of derivatives with respect to the dummy spinors.

\subsection{Massive spinor helicity variables}

Like the massless case, the massive Dirac operator \eqref{eq:pslash} can be factorized over the massive little group $\SO(4)$
by introducing massive spinor helicity variables
\begin{align} \label{eq:m-Id}
    p_A^{\enspace B} 
    &= \vSpinor_A \vert p_{\dot{\alpha}} ]
     [ p^{\dot{\alpha}} \vert^B
    + \vSpinor_A \vert p_{\alpha} \ra\
     \la p^{\alpha} \vert^B
    \,,
\end{align} 
where we sum over repeated massive little group indices $\alpha,\dot{\alpha} = 1,2$. 

Massive spinor helicity (right-bracket) variables
${}_A\vert p_\alpha ]$ and ${}_A\vert p_{\dot{\alpha}} \ra$
satisfy the massive Dirac equations
\begin{align}
    p_A^{\enspace B} 
    \ \vSpinor_B \vert p_{\alpha} \ra
    = i m\  \vSpinor_A \vert p_{\alpha} \ra
    \quad\mathrm{and}\quad 
    p_A^{\enspace B} 
    \ \vSpinor_B \vert p_{\dot{\alpha}} ]
    = - i m\  \vSpinor_A \vert p_{\dot{\alpha}} ]
    \,,
\end{align}
where we assume that $p$ is time-like: $\sqrt{p^2} = \sqrt{-m^2} = \pm i m$. 
Massive spinors with positive imaginary eigenvalues carry Weyl little group indices $\alpha$, and those with negative imaginary eigenvalues carry anti-Weyl little group indices $\dot{\alpha}$.

We use the following representation for the massive spinors
\begin{align} \label{eq:m-spinors}
    \begin{aligned}
    \vSpinor_{A}|p_{\alpha}\rangle 
    = \frac{1}{\sqrt{2}} \begin{pmatrix}
        i (p_2 - i p_3) & \frac{i m - p_4}{p_0 + p_1} \\
        i (p_0 + p_1) & 0 \\
        0 & 1 \\
        m - i p_4 & \frac{- p_2 - i p_3}{p_0 + p_1}
    \end{pmatrix} 
    \,, \quad
    \vSpinor_{A}|p_{\dot\alpha}\rbrack 
    = \frac{1}{\sqrt{2}} \begin{pmatrix}
        i (p_2 - i p_3) & \frac{-i m - p_4}{p_0 + p_1} \\
        i (p_0 + p_1) & 0 \\
        0 & 1 \\
        -m - i p_4 & \frac{- p_2 - i p_3}{p_0 + p_1}
    \end{pmatrix}
    \end{aligned}
\end{align}

The corresponding left brackets are defined by contracting with the charge conjugation matrix \eqref{cmatrix}
\begin{align}
    [p_{\dot{\alpha}}\vert^A 
    = \vSpinor_{B} \vert p_{\dot{\alpha}} ]\ C^{BA}
    \qquad\mathrm{and}\qquad
    \la p_{\alpha}\vert^A = \vSpinor_{B} \vert p_{\alpha} \ra\ C^{BA}
    .
\end{align}
The little group indices $\alpha=1,2$ and $\dot{\alpha}=1,2$ are raised and lower with the charge conjugation matrix
\begin{align}
    C^{\alpha\beta} = \varepsilon^{\alpha\beta} , ~~~ C^{\dot{\alpha}\dot{\beta}} = \varepsilon^{\dot{\alpha}\dot{\beta}} , ~~~ \epsilon^{12} = \epsilon^{\dot{1} \dot{2}} = 1 
\end{align}
as follows
\begin{equation}
    \la p^{\alpha} \vert^B=\la p_{\beta} \vert^B C^{\beta \alpha} , ~~~ \la p^{\dot\alpha} \vert^B=\la p_{\dot\beta} \vert^B C^{\dot\beta \dot\alpha}  .
\end{equation}
The massive brackets also satisfy the completeness relation
\begin{align} \label{eq:m-comprel}
    -i m\  \delta_A^{\enspace B} 
    &= \vSpinor_A \vert p_{\dot{\alpha}} ]\
     [ p^{\dot{\alpha}} \vert^B
    - \vSpinor_A \vert p_{\alpha} \ra\
     \la p^{\alpha} \vert^B
    \,.
\end{align} 

With these conventions, spinor products are computed by contracting the spacetime-Dirac indices
\begin{align}
    \ldirac p_\bullet \vert q_\bullet \rdirac
    = {\ldirac p_\bullet \vert}^A \vSpinor_A\vert q_\bullet \rdirac
    = {\ldirac p_\bullet \vert}_{A}\ C^{AB}\ \vSpinor_B\vert q_\bullet \rdirac
\end{align}
where $\rdirac$ and $\ldirac$ represent either square or angle brackets and the $\bullet$'s are placeholders for either massive ($\alpha,\dot{\alpha},\beta,\dot{\beta},\dots$) or massless ($I,J,\dots$) little group indices.
Like the massless brackets, these spinor products are anti-symmetric
\begin{equation}
    \ldirac p_\bullet \vert q_{\bullet^\prime} \rdirac = - \ldirac q_{\bullet^\prime} \vert p_\bullet \rdirac. 
\end{equation}
Like in the massless case, we use the usual short-hand 
\begin{equation}
  \vSpinor_A\vert i_\alpha \ra = \vSpinor_A\vert p_{i\alpha} \ra
  ~~~~~ \vSpinor_A\vert i_{\dot\alpha} ] = \vSpinor_A\vert p_{i\dot{\alpha}} ]   
\end{equation} to denote the Weyl and anti-Weyl spinors for the $i^\text{th}$ particle in order to simplify formulas.

As in equation \eqref{eq:ul0}, we introduce an underlined bracket for massive particles
\begin{align}
        \vSpinor_A\vert \bul{i} \ra 
        = \vSpinor_A\vert p_{i\alpha} \ra \ z_i^\alpha \,,
        \qquad
        \vSpinor_A\vert \bul{i} ] 
        = \vSpinor_A\vert p_{i\dot{\alpha}} ] \
        \tilde{z}_i^{\dot{\alpha}} \,
\end{align}
to simplify keeping track of little group indices.

Like massless representations, massive representations in five dimensions can be constructed from little group tensors that are fully symmetric in their Weyl $\alpha$ and anti-Weyl  $\dot\alpha$ indices. 
Thus, contracting all little group indices with dummy spinors automatically imposes the correct symmetry properties, and amplitudes become polynomials in the underlined brackets, which are built by contracting spacetime Dirac indices
\begin{align}
    \ldirac \bul{ik} \rdirac
    = {\ldirac \bul{i} \vert}_A\  C^{AB}\ \vSpinor_B\vert \bul{k} \rdirac
    \,, 
    \quad
    \ldirac \bul{i} \ul{l} \ra
    = {\ldirac \bul{i} \vert}_A\  C^{AB}\ \vSpinor_B\vert \ul{l} \ra
    \,, 
    \quad
    \text{and so on.}
\end{align}

\section{Three-point amplitudes \label{sec:3ptAmp}}

In this section, we construct all three-point amplitudes from purely kinematic considerations for any spin and the following configurations of masses: all masses generic and non-vanishing (section \ref{sec:genM}), two massive and distinct with one massless (section \ref{sec:2M10}), two massive and equal with one massless (section \ref{sec:equalmass}), one massive and two massless (section \ref{sec:1M20}), and, all massless (section \ref{sec:3massless}).
In constructing all three-point amplitudes, we assume that the momenta are complex and satisfy momentum conservation $p_1+p_2+p_3=0$.


\subsection{Three massive: unequal masses \label{sec:genM}}

In this section, we construct the space of three-point amplitudes with total spins $S_{i=1,2,3} = s_i + \bar{s}_i$ and masses all non-zero and distinct 
\begin{equation}\label{eq:ampmmm5d}
        A(\mbf{1}^{S_1}, \mbf{2}^{S_2}, \mbf{3}^{S_3}) = \left\{
            \begin{aligned}
                & \ldirac \bul{12}                         \rdirac^{S_1{+}S_2{-}S_3} 
                    \ldirac \bul{23} \rdirac^{S_2{+}S_3{-}S_1} 
                    \ldirac \bul{31} \rdirac^{S_3{+}S_1{-}S_2} 
                && 
                S_1{+}S_2{\geq} S_3 {\geq} |S_1-S_2| 
                \,, \\
                & \ldirac \bul{12} \rdirac^{2S_1} 
                    \ldirac \bul{23} \rdirac^{2S_3} 
                    \la \bul2 | 3 | \bul2 ]^{S_2-S_1-S_3} 
                && 
                S_2 > S_1+S_3 
                \,, \\
                & \ldirac \bul{12} \rdirac^{2S_2} 
                    \ldirac \bul{31} \rdirac^{2S_3} 
                    \la \bul1 | 2 | \bul1 ]^{S_1-S_2-S_3} 
                && 
                S_1 > S_2+S_3 
                \,, \\
                & \ldirac \bul{23} \rdirac^{2S_2} 
                    \ldirac \bul{31} \rdirac^{2S_1} 
                    \la \bul3 | 1 | \bul3 ]^{S_3-S_1-S_2} 
                && 
                S_3 > S_1+S_2 
                \,.
            \end{aligned} \right.
\end{equation}
Here, $s_i,\bar{s}_i$ are half integers denoting the number of chiral and anti-chiral spinors in the amplitude; each monomial contains $2s_i$ $|\bul{i}\rangle$ spinors and  $2\bar{s}_i$ $|\bul{i}]$ spinors.
Since the spinor brackets satisfy non-linear Schouten identities (e.g., \eqref{eq:3massive2}), the amplitude splits into four ``large'' sectors depending on the values for the total spin, $S_i$.
Here, the bracket $\ldirac \bul{kl}\rdirac^{x}$ corresponds to the monomials $\la\bul{kl}\ra^{v_1} \la\bul{kl}]^{v_2} [\bul{kl}\ra^{v_3} [\bul{kl}]^{v_4}$ whose exponents satisfy $v_1 + v_2 +v_3 + v_4 = x$.
When $S_1{+}S_2{\geq} S_3 {\geq} |S_1-S_2|$, there are further subsectors due to cubic identities such as \eqref{eq:massive3idI} and \eqref{eq:futherId}. 
Once a sector with no further relations has been identified, the amplitude is given by any linear combination of monomials whose exponents lie in a polytope dictated by the $s_i$ and $\bar{s}_i$.

Our strategy for building all three-point amplitudes starts by enumerating all possible tensor structures graded by mass dimension. 
We study identities between these tensors at each mass dimension and isolate a minimal spanning set.
At some critical mass dimension, we find that all tensors can be built out of tensors of lower mass dimension. 

\paragraph{The ansatz.}
At mass dimension zero, the possible tensor structures consist only of the charge conjugation matrices.
However, these are anti-symmetric and not permitted in the amplitude. 
Enumerating all possible structures with mass dimension one reveals the following candidates 
\begin{align} \label{eq:unequal2Brackets}
    \ldirac \bul{12} \rdirac
    \,,\quad 
    \ldirac \bul{13} \rdirac 
    \,,\quad
    \ldirac \bul{23} \rdirac
    \,.
\end{align}
All other mass dimension one tensors vanish or are linearly related to these. 

At mass dimension two, the candidate tensors are: $\ldirac\bul{i}\vert j \vert\bul{k}\rdirac$. 
Obviously, we do not have to consider tensors with $i=j=k$ since these vanish. 
Moreover, we must have $i = k \neq j$. Otherwise, the tensor is reducible to a mass dimension one two-bracket \eqref{eq:unequal2Brackets} via the Dirac equation. 
Then, since the brackets $\la \bul{i}\vert j \vert\bul{i}\ra = [\bul{i}\vert j \vert\bul{i}] = 0$, we are left with the following candidate tensors
\begin{align}
    \langle \bul 1 | 2 | \bul 1 \rbrack
    \,, \quad 
    \langle \bul 2 | 3 | \bul 2 \rbrack
    \,, \quad 
    \langle \bul 3 | 1 | \bul 3 \rbrack
    \,.
\end{align}
Note that we do not have to consider $\la\bul{1}\vert3\vert\bul{1}]$ (and others like it) because it is equivalent to $\la\bul{1}\vert2\vert\bul{1}]$ up to lower mass dimension tensors. 
Furthermore, the reflections of the above brackets (e.g., $[\bul 1 | 2 | \bul 1 \ra$) do not need to be considered since $[\bul i | j | \bul i \ra = \la \bul i | j | \bul i ]$.  

Thus, all tensor structures up to mass dimension two can be built from linear combinations of the following building blocks
\begin{equation}\label{eq:setmmm5d}
        \begin{aligned}
            & 
            \langle \bul{12} \rangle
            \,, \quad 
            \lbrack \bul{12} \rangle
            \,, \quad 
            \langle \bul{12} \rbrack
            \,, \quad 
            \lbrack \bul{12} \rbrack
            \,, \quad
            \langle \bul{31} \rangle
            \,, \quad 
            \lbrack \bul{31} \rangle
            \,, \quad 
            \langle \bul{31} \rbrack
            \,, \quad 
            \lbrack \bul{31} \rbrack
            \,, 
            \\
            & 
            \langle \bul{23} \rangle
            \,, \quad 
            \lbrack \bul{23} \rangle
            \,, \quad 
            \langle \bul{23} \rbrack
            \,, \quad 
            \lbrack \bul{23} \rbrack
            \,, \quad
            \langle \bul1 | 2 | \bul1 \rbrack
            \,, \quad 
            \langle\bul 2 | 3 | \bul2 \rbrack
            \,, \quad 
            \langle \bul3 | 1 | \bul3 \rbrack
            \,.
        \end{aligned}
\end{equation}
It also turns out that mass dimension three and higher tensors can all be built from products of \eqref{eq:setmmm5d}.
For example, $\ldirac \bul 1 \vert 2 \vert 3 \vert \bul 2 \rdirac = 2 p_2 \cdot p_3 \ldirac \bul 1\vert \bul 2 \rdirac \pm i m_2 \ldirac \bul 1 \vert 3 \vert \bul 2 \rdirac$ and all others like it are reducible to lower mass dimension tensors.

So far, we have only used linear identities between tensor structures to constrain the list of allowed tensor structures. 
Generically, there will be quadratic Schouten identities involving the tensors in \eqref{eq:setmmm5d}
\begin{equation}\label{eq:3massive2}
    \begin{aligned}
        \langle \bul{1 2} \rangle 
            \, \langle \bul{3} | 1 | \bul{3} \rbrack 
        &= \langle \bul{1 3} \rbrack 
            \langle \bul{3} | 1 | \bul{2} \rangle 
        - \langle \bul{1 3} \rangle 
            \lbrack \bul{3} | 1 | \bul{2} \rangle 
        + m_1 \langle \bul{3 1} \rangle \langle \bul{2 3} \rbrack 
        + m_1 \lbrack \bul{3 1} \rangle \langle \bul{2 3} \rangle 
        \\ 
        &= (m_1-m_2+m_3) \,
        \langle \bul{1 3} \rbrack \,
        \langle \bul{3 2} \rangle
        + (m_1-m_2-m_3) \,
        \langle \bul{1 3} \rangle \,
        \lbrack \bul{3 2} \rangle \,,
    \end{aligned}
\end{equation}
along with the permutations of particles 1, 2, and 3 and exchanging angle and square brackets. 
In particular, \eqref{eq:3massive2} implies that the tensors on the left-hand side above cannot appear together.\footnote{Mathematically, this means that we are working in the polynomial quotient ring 
\begin{align*}
    R=
    \frac{\mathbb{C}\big[
        \eqref{eq:setmmm5d}
    \big]}{\la 
        \eqref{eq:3massive2}, \text{variations}\eqref{eq:3massive2}
    \ra}\,. 
\end{align*}
Since there are often many non-linear identities between spinor helicity variables, it is important to understand the ring structure when constructing ansatz. See \cite{DeLaurentis:2022otd, DeLaurentis:2022knk} for more on the connections of spinor helicity variables to algebraic geometry and $p$-adic numbers.
} 
By counting the total powers of spinors, we arrive at the ansatz \eqref{eq:ampmmm5d} for five-dimensional unequal massive amplitudes where \eqref{eq:3massive2} splits the ansatz into four sectors.
There are further cubic identities (arising from the composition of various Schouten identities) that split the sector $S_1{+}S_2{\geq} S_3 {\geq} |S_1-S_2|$  into further sub-sectors. 

In the remainder of this section, we will detail how the cubic identities affect the ansatz \eqref{eq:ampmmm5d} and the counting of the number of independent amplitudes. 
In particular, we find the expected number of independent amplitudes \cite{Costa:2011mg, Elkhidir:2014woa}. 

\paragraph{Case 1: $\boldsymbol{S_1+S_2} \boldsymbol\geqslant \boldsymbol{S_3} \boldsymbol\geqslant \boldsymbol{|S_1-S_2|}$.} 
For the first case, the explicit ansatz is
\begin{equation}
\begin{aligned}
    A(\mbf{1}^{s_1+\bar{s}_1}, \mbf{2}^{s_2+\bar{s}_2}, \mbf{3}^{s_3 + \bar{s}_3}) &=
    \langle \bul{12} \rangle^{v_1} \langle \bul{12} \rbrack^{v_2} 
        \langle \bul{13} \rangle^{v_3} \langle \bul{13} \rbrack^{v_4} 
        \lbrack \bul{12} \rangle^{v_5} \lbrack \bul{12} \rbrack^{v_6} 
    \\&\qquad\times 
    \lbrack \bul{13} \rangle^{v_7} \lbrack \bul{13} \rbrack^{v_8} 
        \langle \bul{23} \rangle^{v_9} \langle \bul{23} \rbrack^{v_{10}} 
        \lbrack \bul{23} \rangle^{v_{11}} \lbrack \bul{23} \rbrack^{v_{12}}
\end{aligned}
\end{equation}
with the exponents satisfying
\begin{equation} \label{eq:expConstraints}
    \begin{aligned} 
        v_1 + v_2 + v_3 + v_4 = 2 s_1 
        \,, & \qquad
        v_5 + v_6 + v_7 + v_8 = 2 \bar{s}_1 
        \,, \\
        v_1 + v_5 + v_9 + v_{10} = 2 s_2 
        \,, & \qquad
        v_2 + v_6 + v_{11} + v_{12} = 2 \bar{s}_2 
        \,, \\
        v_3 + v_7 + v_{9} + v_{11} = 2 s_{3} 
        \,, & \qquad
        v_{4} + v_{8} + v_{10} + v_{12} = 2 \bar{s}_3 
        \,,
    \end{aligned}
\end{equation}
and $v_i \in \mathbb{N}_0$.
Therefore, the final solution space of independent amplitudes is seemingly given by the lattice points inside a $(12 - 6) = 6$ dimensional polytope cut out by the above conditions.

However, there are further cubic identities satisfied by these brackets that come from the composition of \eqref{eq:3massive2} and its variations
\begin{equation}\label{eq:massive3idI}
    \begin{aligned}
        0&=
        {\color{violet} \la \bul{12} ] } 
            {\color{violet} [ \bul{13} \ra } 
            {\color{violet} \la \bul{23} ] }
        - [ \bul{23} \ra \la \bul{13} ] [ \bul{12} \ra
        - \frac{m_{++}}{m_{--}} 
            {\color{violet} \la \bul{12}\ra } [ \bul{23} \ra {\color{violet} [\bul{13} ] }
        + \frac{m_{++}}{m_{--}} 
            {\color{BrickRed} 
                \langle \bul{13} \rangle 
                \lbrack \bul{12} \rbrack 
                \langle \bul{23} \rbrack 
            }
        \,,
        \\
        0&=
        {\color{violet} \la \bul{12} ] } {\color{violet} [ \bul{13} \ra } {\color{violet} \la \bul{23} ] }
        - [ \bul{23} \ra \la \bul{13} ] [ \bul{12} \ra
        - \frac{m_{++}}{m_{-+}} 
            {\color{BrickRed} 
                \la \bul{12} \ra
                [ \bul{13} \ra 
                [ \bul{23} ] 
            }
        + \frac{m_{++}}{m_{-+}} 
            {\color{violet} \la \bul{23} \ra } \la \bul{13} ] {\color{violet} [ \bul{12} ] }
        \,,
        \\
        0&=
        {\color{violet} \la \bul{12} ] } 
            {\color{violet} [ \bul{13} \ra } 
            {\color{violet} \la \bul{23} ] }
        - [ \bul{23} \ra 
            \la \bul{13} ] 
            [ \bul{12} \ra
        + \frac{m_{++}}{m_{+-}} 
            {\color{violet} \la \bul{13} \ra } 
            [ \bul{12} \ra 
            {\color{violet} [ \bul{23} ] }
        - \frac{m_{++}}{m_{+-}} 
            {\color{BrickRed} 
                \langle \bul{12} \rbrack 
                \lbrack \bul{13} \rbrack 
                \langle \bul{23} \rangle 
            }
        \,,
    \end{aligned}
\end{equation}
where $m_{\pm\pm} = m_1 \pm m_2 \pm m_3$.
Using (\ref{eq:massive3idI}), we can eliminate the monomials $\langle \bul{13} \rangle \lbrack \bul{12} \rbrack \langle \bul{23} \rbrack$, $\langle \bul{12} \rangle \lbrack \bul{13} \rangle \lbrack \bul{23} \rbrack$, and $\langle \bul{12} \rbrack \lbrack \bul{13} \rbrack \langle \bul{23} \rangle$ (colored in {\color{BrickRed} red}) in favor of the other monomials. 
The {\color{violet}purple} brackets indicate a repeated two-bracket from the set of {\color{BrickRed}red} monomials that were eliminated.
This means that at least one exponent in each list $(v_3, v_{6}, v_{10})$, $(v_1, v_7, v_{12})$, $(v_{2}, v_{8}, v_9)$ must be zero. Therefore, we obtain $3^3 = 27$ sub-sectors. Each of them has three exponents fixed to zero, so the dimension of each sector becomes $(6-3) = 3$. 

However, nothing prevents there from being additional identities \textit{inside} in each of these 27 sectors. 
There are further identities in the sectors where the following monomials were eliminated
\begin{equation}\label{eq:subsectors}
    \begin{aligned}
        \langle \bul{12} \rangle \langle \bul{13} \rangle \langle \bul{12} \rbrack \,, \quad \langle \bul{13} \rbrack \langle \bul{13} \rangle \langle \bul{23} \rangle \,, \\
        \langle \bul{23} \rangle \langle \bul{12} \rangle \langle \bul{23} \rbrack \,, \quad \langle \bul{12} \rbrack \langle \bul{12} \rangle \langle \bul{13} \rangle \,, \\
        \langle \bul{13} \rangle \langle \bul{23} \rangle \langle \bul{13} \rbrack \,, \quad \langle \bul{23} \rbrack \langle \bul{23} \rangle \langle \bul{12} \rangle \,.
    \end{aligned}
\end{equation}
For example, the additional identity in the sub-sector where $\langle \bul{12} \rangle \langle \bul{31} \rangle \langle \bul{12} \rbrack$ was eliminated is
\begin{equation} \label{eq:futherId}
    \begin{aligned}
        0 &= m_{++} \lbrack\bul{12}] 
            \langle\bul{13}]  \lbrack\bul{23}\rangle 
        - m_{+-} \lbrack\bul{12}] 
            \lbrack\bul{13}\rangle  \langle\bul{23}]
        + m_{--} \lbrack\bul{12}\rangle 
            \lbrack\bul{13}\rangle [\bul{23}] 
        - m_{-+} \lbrack\bul{12}\rangle 
            \lbrack\bul{13}] [\bul{23}\rangle
        \,.
    \end{aligned}
\end{equation}
The identities in other sectors in \eqref{eq:subsectors} can be derived from chiral flips and cyclic permutations of \eqref{eq:massive3idI}. 
Such identities reduce the previously three-dimensional solution space to three two-dimensional boundaries.

\paragraph{Case 2: $\boldsymbol{S_2>S_1+S_3}$.} 
There are no further identities for the last three cases in (\ref{eq:ampmmm5d}). 
In the second case, the explicit ansatz reads
\begin{equation}\begin{aligned}
    A(\mbf{1}^{s_1 + \bar{s}_1}, \mbf{2}^{s_2 + \bar{s}_2}, \mbf{3}^{s_3 + \bar{s}_3}) &= 
    \langle \bul{12} \rangle^{w_1} 
    \langle \bul{12} \rbrack^{w_2} 
    \lbrack \bul{12} \rangle^{w_3} 
    \lbrack \bul{12} \rbrack^{w_4} 
    \langle \bul{23} \rangle^{w_5} 
    \\&\qquad\times 
    \lbrack \bul{23} \rangle^{w_6} 
    \langle \bul{23} \rbrack^{w_7} 
    \lbrack \bul{23} \rbrack^{w_8} 
    \langle \bul2 | 3 | \bul2 ]^{S_2-S_1-S_3} \,,
 \end{aligned}\end{equation}
with the exponents satisfying
\begin{equation}
    \begin{aligned}
        w_1 + w_2 = 2 s_1 
        \,, &\quad
        w_3 + w_4 = 2 \bar{s}_1 
        \,, \\
        w_5 + w_6 = 2 s_{3} 
        \,, &\quad
        w_7 + w_8 = 2 \bar{s}_3 
        \,, \\
        w_1 + w_3 + w_5 + w_7 &= s_2 - \bar{s}_2 + S_1 + S_3 
        \,,
    \end{aligned}
\end{equation}
and $w_i \in \mathbb{N}_0$.
Since the exponent for the final tensor $\langle \bul2 | 3 | \bul2 \rbrack$ is fixed,  the final solution space has $8-5=3$ dimensions. 
 Choosing $(w_1,w_3,w_5)$ to parametrize this space, we obtain a polytope with faces defined by the following conditions:
\begin{equation}
    \begin{aligned}
        &
        0 \leqslant w_1 \leqslant 2s_1 
        \,, \qquad
        0 \leqslant w_3 
        \leqslant 2\bar{s}_1 
        \,, \qquad
        0 \leqslant w_5 \leqslant 2s_3 
        \,, \\ &
        S_1 + s_2 - \bar{s}_2 + s_3 - \bar{s}_3 
        \leqslant w_1+w_3+w_5 
        \leqslant s_2 - \bar{s}_2 + S_1 + S_3 
        \,.
    \end{aligned}
\end{equation}
The number of independent amplitudes for the second line in (\ref{eq:ampmmm5d}) is given by the number of lattice points inside or on the boundary of this polytope.

\paragraph{All cases.} 
Similar to the $S_2>S_1+S_3$ sector, one can construct the polytope and amplitudes in the third and fourth sectors of (\ref{eq:ampmmm5d}). 
In the end, the total polytope that counts the number of independent three-point massive amplitudes and couplings is the direct sum of the 27 subsectors of case 1 as well as three polytopes from cases 2, 3, and 4.

\paragraph{Comparing to known results for the number of independent amplitudes.}
Above, we outlined how to compute the number of independent amplitudes by quotienting out by the identities \eqref{eq:massive3idI} and \eqref{eq:futherId}. 
In particular, obtaining explicit solutions for low spin ($S\leq4$) in \texttt{Mathematica} is not too difficult. 
We have compared the output of the above counting strategy to known closed-form results from \cite{Costa:2011mg} and \cite{Elkhidir:2014woa} and found agreement. 
We will be more explicit about this construction and counting in the next section, which has fewer sectors.

\subsection{Two massive and one massless: unequal masses\label{sec:2M10}}

In this and the next section, we set $m_3=0$ and assume that $m_{i=1,2}>0$. In this section, we set $m_1 \neq m_2$, while in the next section, we will take $m_1 = m_2$.

As before, we make an ansatz for the amplitude compatible with little group scaling from a list of tensors that are not linearly related and then impose any non-linear relations satisfied by these tensors. The amplitude splits into three ``large'' sectors depending on the values of spin $s_1,s_2,\bar{s}_1,\bar{s}_2$ and the helicity $S_3$
\begin{align}\label{2 massive unequal}
    A(\mbf{1}^{s_1{+}\bar{s}_1}, \mbf{2}^{s_2{+}\bar{s}_2}, 3^{S_3})
    =
    \begin{cases}
        \ldirac \bul{12} \rdirac ^{S_1{+}S_2{-}S_3}\ldirac \bul{2}\ul{3} \rangle^{S_2-S_1+S_3}\ldirac \bul{1} \ul{3}\rangle^{S_1-S_2+S_3} 
        & S_1 {+} S_2 {\geqslant} S_3 \geqslant |S_1{-}S_2|,
        \\
        \ldirac \bul{12} \rdirac ^{2S_2}\ldirac \bul{2}\ul{3}\rangle^{S_3}\langle \bul{2}|3|\bul{2} \rbrack^{S_2{-}S_1{-}S_3} 
        & S_2 \geqslant S_1 {+} S_3 \geqslant |s_2{-}\bar{s}_2|,
        \\
        \ldirac \bul{12} \rdirac^{2S_2}\langle \bul{1}|3|\bul{1} \rbrack ^{S_1-S_2-S_3}\ldirac \bul{1} \ul{3} \rangle^{2S_3} 
        & S_1 \geqslant S_2 {+} S_3 \geqslant |s_1{-}\bar{s}_1|.
    \end{cases}
\end{align}
As before, the bracket $\ldirac \bul{kl}\rdirac^{x}$ corresponds to the monomials $\la\bul{kl}\ra^{v_1} \la\bul{kl}]^{v_2} [\bul{kl}\ra^{v_3} [\bul{kl}]^{v_4}$ whose exponents satisfy $v_1 + v_2 +v_3 + v_4 = x$. 

\paragraph{The ansatz.}
Like in section \ref{sec:genM}, we list all possible mass dimension one and two tensors, impose linear relations, and discard any mass dimension two tensors that can be reduced to mass dimension one tensor via momentum conservation or the Dirac equation. 
In the end, we find that the following  set of tensor structures  
\begin{equation}\label{eq:5dset0mm}
        \langle \ul{3} \bul{1} \rangle 
        \,,\, 
        \langle \ul{3} \bul{1} \rbrack 
        \,,\, 
        \langle \ul{3} \bul{2} \rangle 
        \,,\, 
        \langle \ul{3} \bul{2} \rbrack 
        \,,\, 
        \langle \bul{1} \bul{2} \rangle 
        \,,\, 
        \langle \bul{1} \bul{2} \rbrack 
        \,,\, 
        \lbrack \bul{1} \bul{2} \rangle 
        \,,\, 
        \lbrack \bul{1} \bul{2} \rbrack 
        \,,\, 
        \langle \bul{1} | 3 | \bul{1} \rbrack 
        \,,\, 
        \langle \bul{2} | 3 | \bul{2} \rbrack 
        \,,
\end{equation}
is enough to span all higher mass dimension tensors. 

Thus, the general ansatz for these amplitudes is
\begin{equation}
\begin{aligned}
    A(\mbf{1}^{s_1{+}\bar{s}_1}, \mbf{2}^{s_2{+}\bar{s}_2}, 3^{S_3})
    &= 
    \langle \ul{3} \bul{1} \rangle^{v_1}
        \langle \ul{3} \bul{1} \rbrack^{v_2} 
        \langle \ul{3} \bul{2} \rangle^{v_3} 
        \langle \ul{3} \bul{2} \rbrack^{v_4} 
        \langle \bul{1} \bul{2} \rangle^{v_5} 
        \\&\quad\times 
        \langle \bul{1} \bul{2} \rbrack^{v_6} 
        \lbrack \bul{1} \bul{2} \rangle^{v_7} 
        \lbrack \bul{1} \bul{2} \rbrack^{v_8} 
        \langle \bul{1} | 3 | \bul{1} \rbrack^{v_9} 
        \langle \bul{2} | 3 | \bul{2} \rbrack^{v_{10}} 
        \,,
\end{aligned}
\end{equation}
where the exponents $v_i$ are subject to the following constraints 
\begin{align}
\label{eq:2M10_vCon}
\begin{aligned}
 v_1+v_2+v_3+v_4 = 2 S_3 \,,
 & \qquad
 v_1+v_5+v_6+v_9 = 2 s_1  \,,
 \\
 v_3+v_5+v_7+v_{10} = 2 s_2 \,,
 &\qquad 
 v_2+v_7+v_8+v_9 = 2 \bar{s}_1 \,,
 \\
 v_4+v_6+v_8+v_{10} = 2 \bar{s}_2 \,,
 &\qquad
v_i \in \mathbb{N}_0 \,,
\end{aligned}
\end{align}
which ensure that there are $s_a$ spinors $\vert \bul{a}\ra$, $\bar{s}_{a}$ spinors $\vert \bul{a}]$ for $a=1,2$ and $S_3$ spinors $\vert \ul{3} \ra$.

There are also quadratic identities involving the tensor structures \eqref{eq:5dset0mm} that further constrain the exponents $v_i$. 
For example, 
\begin{align}\label{eq:tensorRelation-102M1}
    \langle \bul{1} | 3 | \bul{1} \rbrack \langle \bul{2} \ul{3} \rangle 
    = \langle \bul{2} | 3 | \bul{1} \rbrack \langle \bul{1} \ul{3} \rangle 
    + \langle \bul{1} | 3 | \bul{2} \rangle \lbrack \bul{1} \ul{3} \rangle 
    = i \left(
        m_- \langle \bul{2} \bul{1} \rbrack  
            \langle \bul{1} 3 \rangle
        - m_+ \langle \bul{1} \bul{2} \rangle 
            \lbrack \bul{1} \ul{3} \rangle 
    \right)
\end{align}
where $m_\pm = m_1 \pm m_2$.
Similar, identities exist when $1 \leftrightarrow 2$ and $|\bul{i}\rangle \leftrightarrow |\bul{i}\rbrack$ for $i=\{1,2\}$. We also find another identity
\begin{equation}\begin{aligned}\label{eq:tensorRelation-102M2}
    \langle \bul{1} | 3 | \bul{1} \rbrack \lbrack \bul{2} | 3 | \bul{2} \rangle 
    = \langle \bul{1} | 3 | \bul{2} \rangle \lbrack \bul{2} | 3 | \bul{1} \rbrack - \langle \bul{1} | 3 | \bul{2} \rbrack \langle \bul{2} | 3 | \bul{1} \rbrack 
    = -m_+^2 \langle \bul{1} \bul{2} \rangle \lbrack \bul{1} \bul{2} \rbrack - m_-^2 \langle \bul{1} \bul{2} \rbrack \langle \bul{2} \bul{1} \rbrack \,.
\end{aligned}\end{equation}
Therefore, the following pairs of tensors cannot appear together in our ansatz 
$(\langle \bul{1} \ul{3} \rangle , \langle \bul{2} | 3 | \bul{2} \rbrack)$, 
$(\lbrack \bul{1} \ul{3} \rangle , \langle \bul{2} | 3 | \bul{2} \rbrack)$, 
$(\langle \bul{2} \ul{3} \rangle , \langle \bul{1} | 3 | \bul{1} \rbrack)$, $(\lbrack \bul{2} \ul{3} \rangle , \langle \bul{1} | 3 | \bul{1} \rbrack)$ 
and  
$(\langle \bul{1} | 3 | \bul{1} \rbrack , \langle \bul{2} | 3 | \bul{2} \rbrack)$. 
In terms of exponents, there needs to be at least one zero in each of the following pairs $(v_1,v_{10})$, $(v_2,v_{10})$, $(v_3,v_{9})$, $(v_4,v_{9})$ and $(v_9,v_{10})$.
This condition breaks the amplitude into the three ``large'' sectors as highlighted in \eqref{2 massive unequal} 
\begin{align}\begin{aligned}
    (v_9=0,v_{10}=0)
    &\implies 
    S_1 {+} S_2 \geqslant S_3 \geqslant |S_1{-}S_2|
    \,,
    \\
    (v_9=0,v_{10}>0) 
    &\implies 
    S_2 \geqslant S_1 {+} S_3 \geqslant |s_2{-}\bar{s}_2|
    \,,
    \\
    (v_9>0,v_{10}=0)
    &\implies
    S_1 \geqslant S_2 {+} S_3 \geqslant |s_1{-}\bar{s}_1|
    \,.
\end{aligned}\end{align}

\begin{figure}
    \centering
    \includegraphics[width=.3\textwidth]{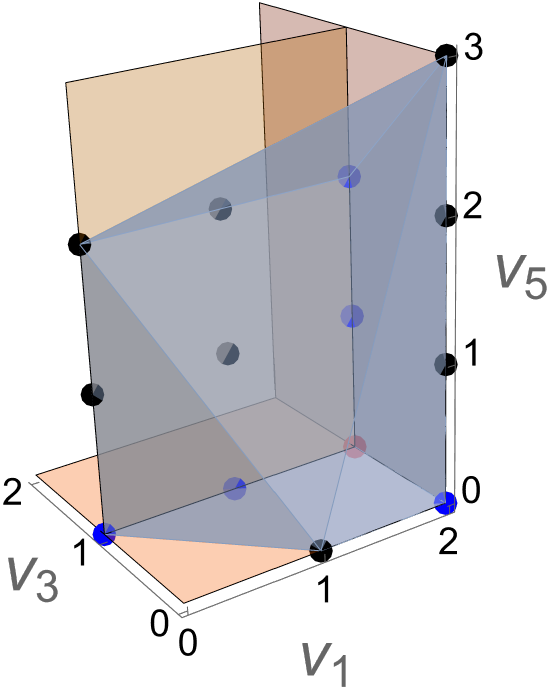}
    \\
    \includegraphics[width=.3\textwidth]{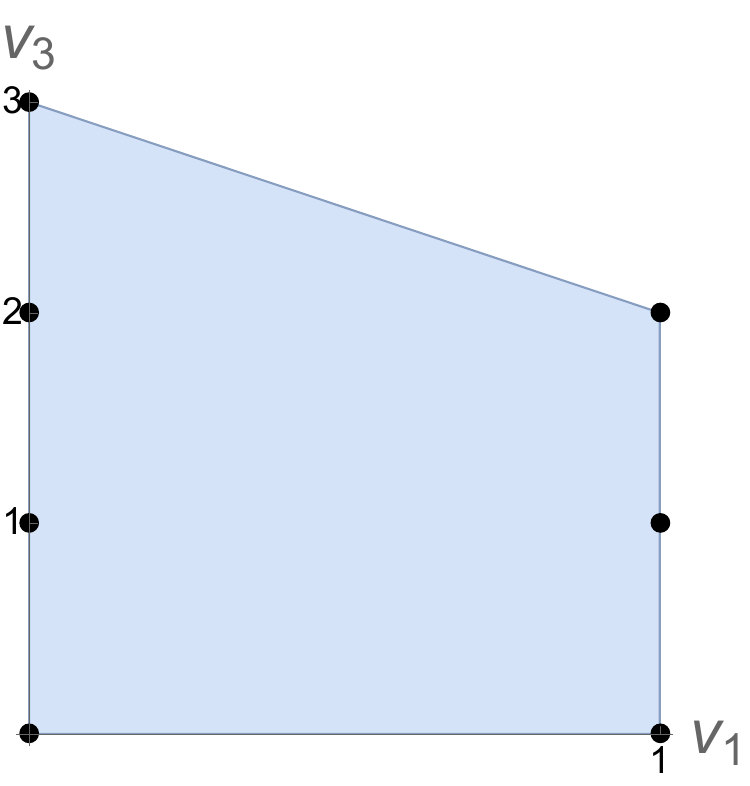}
    \quad
    \includegraphics[width=.3\textwidth]{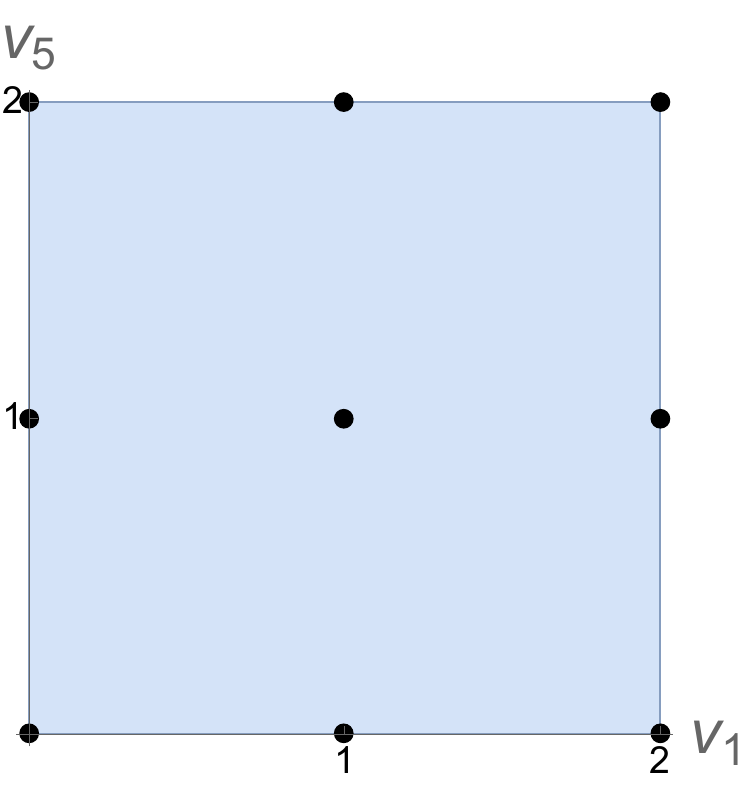}
    \quad
    \includegraphics[width=.3\textwidth]{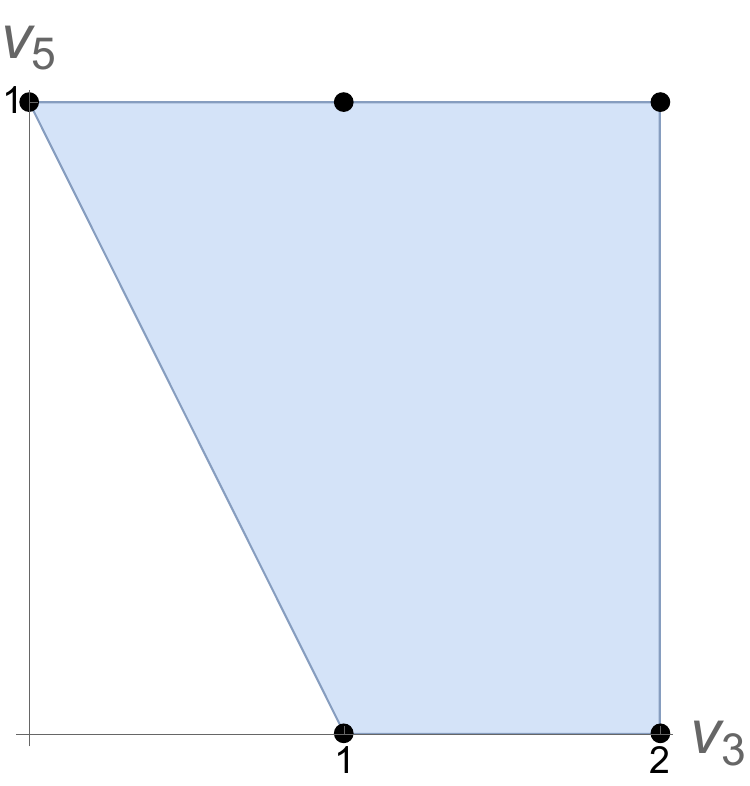}
    \caption{The polytope $P_{00}$ for $(s_1,\bar{s}_1,s_2,\bar{s}_2,S_3) = (\frac{5}{2}, 2, \frac{3}{2}, \frac{5}{2}, \frac{3}{2})$. 
    However, generally, only the lattice points on the planes defined by $v_2=0$, $v_4=0$, and $v_5=0$ (depicted in orange above) are independent.
    The intersection of $P_{00}$ with the planes $v_2=0$, $v_4=0$, and $v_5=0$ are also depicted in the lower portion of this figure.
    For this spin configuration, there are a total of 14 independent amplitudes in the $P_{00}$ sector.
    The allowed point in the triple intersection of the {\color{Orange}orange} planes is denoted in {\color{red}red} while the allowed points in the double intersection of the {\color{Orange}orange} planes are denoted in {\color{blue}blue}. 
    The {\color{red}red} and {\color{blue}blue} points should only be counted once.
    } 
    \label{fig:00Sol}
\end{figure}

\paragraph{Case 1: $\boldsymbol{S_1 {+} S_2 \geqslant S_3 \geqslant |S_1{-}S_2|}$.}

When $v_9=v_{10}=0$, there are no 3-brackets in the ansatz. 
However, there is the cubic identity 
\begin{align} \label{eq:2M10Cubic}
    -m_{-} \la\ul{3}\bul{1}] \la\ul{3}\bul{2}\ra  \la\bul{12}]
    +m_{+} \la\ul{3}\bul{1}] \la\ul{3}\bul{2}]  \la\bul{12}\ra
    +m_{-} \la\ul{3}\bul{1}\ra \la\ul{3}\bul{2}] [\bul{12}\rangle
    -m_{+} \la\ul{3}\bul{1}\ra \la\ul{3}\bul{2}\ra [\bul{12}]
    = 0
    \,.
\end{align}
that comes from the composition of Schouten identities. 
This means that we can eliminate $\langle \ul{3} \bul{1} \rbrack \langle \ul{3} \bul{2} \rbrack \langle \bul{1} \bul{2} \rangle $ for the other monomials above. Therefore, at least one of the exponents $(v_2, v_4, v_5) = (S_1+S_3-S_2-v_1,S_2+S_3-S_1-v_3,v_5)$ must also be set to zero creating three sectors (see the {\color{orange}orange} planes in figure \ref{fig:00Sol}). 
Note that one could choose to eliminate a different monomial appearing in the cubic identity \eqref{eq:2M10Cubic}, resulting in an equivalent representation for the same amplitudes.

Thus, the allowed amplitudes of this sector are 
\begin{align}\begin{aligned} \label{eq:A00}
    A(\mbf{1}^{s_1{+}\bar{s}_1}, \mbf{2}^{s_2{+}\bar{s}_2}, 3^{S_3})
    &= \sum_{(v_1, v_3, v_5) \in \mathbb{Z}^3 \cap P_{00} \cap \{v_2=0\} } g_{v_1,v_3,v_5} A_{v_1,v_3, v_5}(\mbf{1}^{s_1+\bar{s}_1}, \mbf{2}^{s_2+\bar{s}_2}, 3^{S_3})
    \\&\quad
    + \sum_{(v_1, v_3, v_5) \in \mathbb{Z}^3 \cap P_{00} \cap \{v_4=0\} } g_{v_1,v_3,v_5} A_{v_1,v_3, v_5}(\mbf{1}^{s_1+\bar{s}_1}, \mbf{2}^{s_2+\bar{s}_2}, 3^{S_3})
    \\&\quad
    +\sum_{(v_1, v_3, v_5) \in \mathbb{Z}^3 \cap P_{00} \cap \{v_5=0\} } g_{v_1,v_3,v_5} A_{v_1,v_3, v_5}(\mbf{1}^{s_1+\bar{s}_1}, \mbf{2}^{s_2+\bar{s}_2}, 3^{S_3})
\end{aligned}\end{align}
where we have solved the constraints \eqref{eq:2M10_vCon} leaving $v_1, v_3$ and $v_5$ free and defined
\begin{align} \label{eq:A00-2M10}
    A_{v_1,v_3, v_5}(\mbf{1}^{s_1{+}\bar{s}_1}, \mbf{2}^{s_2{+}\bar{s}_2}, 3^{S_3})&= 
    \langle \ul{3} \bul{1} \rangle^{v_1}
    \langle \ul{3} \bul{1} \rbrack^{S_3+S_1-S_2-v_1} 
    \langle \ul{3} \bul{2} \rangle^{v_3} 
    \langle \ul{3} \bul{2} \rbrack^{S_3+S_2-S_1-v_3}    
    \langle \bul{1} \bul{2} \rangle^{v_5}
    \\&\quad\times
    \langle \bul{1} \bul{2} \rbrack^{2 s_1 - v_1 - v_5}
    \lbrack \bul{1} \bul{2} \rangle^{2 s_2 - v_3 - v_5}
    \lbrack \bul{1} \bul{2} \rbrack^{\bar{s}_1-s_1 + \bar{s}_2-s_2 - S_3 + v_1 + v_3 + v_5}
    \,.
    \nn
\end{align}
Here, $P_{00}$ is
the polyhedron with the defining faces
\begin{align}\label{eq:P00}
P_{00}: \left\{
\begin{matrix*}[l]
    & v_{i=1,3,5} \geq 0, 
    &
    & S_3+S_1-S_2-v_1 \geq 0,
    \\
    & S_3+S_2-S_1-v_3 \geq 0,
    &
    & 2 s_1 - v_1 - v_5 \geq 0, 
    \\
    & 2 s_2 - v_3 - v_5 \geq 0, 
    &
    & \bar{s}_1-s_1 + \bar{s}_2-s_2 - S_3 
        + v_1 + v_3 + v_5 &\geq 0 
    \,,
\end{matrix*}
\right.
\end{align}
and 
\begin{align}
     (
     S_3+S_1-S_2,
     S_3+S_2-S_1,
     \bar{s}_1-s_1 + \bar{s}_2-s_2 - S_3
     ) \in \mathbb{Z}^3
     \,,
\end{align}
for the formula \eqref{eq:A00-2M10} to make sense.
The number of independent amplitudes in this sector is given by the number of lattice points in $P_{00}$ that lie on the planes defined by $v_2=0$, $v_4=0$ and $v_5=0$ (an example is provided in figure \ref{fig:00Sol}).

\paragraph{Case 2: $\boldsymbol{S_1 \geqslant S_2 {+} S_3 \geqslant |s_1{-}\bar{s}_1|}$.}
For $v_9=0$ and $v_{10}>0$, the independent amplitudes are
\begin{align} \label{eq:A01}
    A(\mbf{1}^{s_1{+}\bar{s}_1}, \mbf{2}^{s_2{+}\bar{s}_2}, 3^{S_3}) &= 
    \sum_{(v_1, v_5) \in P_{01} \cap \mathbb{Z}^3} g_{v_1,v_5} A_{v_1, v_5}(\mbf{1}^{s_1+\bar{s}_1}, \mbf{2}^{s_2+\bar{s}_2}, 3^{S_3})
\end{align}
where 
\begin{align}\label{eq:A01-2M10}
\begin{aligned}
    A_{v_1, v_5}(\mbf{1}^{s_1{+}\bar{s}_1},\mbf{2}^{s_2{+}\bar{s}_2}, 3^{S_3}) &= 
        \langle \ul{3} \bul{1} \rangle^{v_1} 
        \langle \ul{3} \bul{1} \rbrack^{2 S_3-v_1}  
        \langle \bul{1} \bul{2} \rangle^{v_5}
        \langle \bul{1} \bul{2} \rbrack^{2 s_1-v_1-v_5}        
        \lbrack \bul{1} \bul{2} \rangle^{S_1 + s_2 - \bar{s}_2 - S_3 - v_5} 
        \\&\quad\times        
        \lbrack \bul{1} \bul{2} \rbrack^{\bar{s}_1 -s_1 + \bar{s}_2 - s_2 - S_3 + v_1 + v_5} 
        \langle \bul{2} | 3 | \bul{2} \rbrack^{S_2 + S_3 - S_1 } , 
\end{aligned}
\end{align}
and $P_{01}$ is the polyhedron with the defining faces
\begin{align}
P_{01}: \left\{
\begin{matrix*}[l]
    & v_{i=1,5} \geq 0, 
    &
    & 2 S_3-v_1 \geq 0,
    \\
    & S_1 + s_2 - \bar{s}_2 - S_3 - v_5 \geq 0,
    &
    & 2 s_1-v_1-v_5 \geq 0, 
    \\
    & \bar{s}_1 -s_1 + \bar{s}_2 - s_2 - S_3 + v_1 + v_5\geq 0 .
    &
    &
\end{matrix*}
\right.
\end{align}
The last face above does not yield any interesting constraint and can be ignored.
The number of independent amplitudes is given by the number of lattice points inside this polyhedron.

\begin{figure}
    \centering
    \includegraphics[scale=.5]{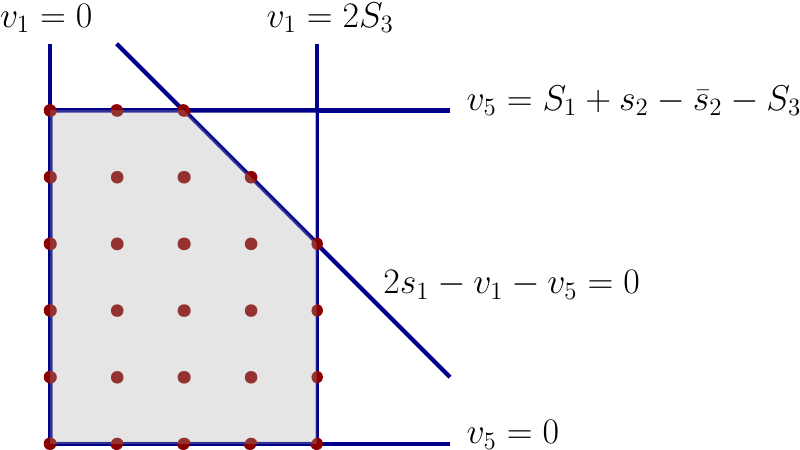}
    \caption{Sketch of $P_{01}$. Depending on the parameters, the line $2 s_1-v_1-v_5 =0$ may or may not intersect the lines $v_1=2S_3$ and $v_5 = S_1 + s_2 - \bar{s}_2 - S_3$ below their intersection point. }
    \label{fig:enter-label}
\end{figure}

\paragraph{Case 3: $\boldsymbol{S_1 \geqslant S_2 {+} S_3 \geqslant |s_1{-}\bar{s}_1|}$. }
The final category of amplitudes has $v_{10}=0$ and $v_{9}>0$. 
These amplitudes can be obtained from the \eqref{eq:A01-2M10} by replacing $v_1\to v_3$ and $\bul{1}\to \bul{2}$.
We checked that the amplitudes given by equations \eqref{eq:A00-2M10} and \eqref{eq:A01-2M10} are linearly independent for all $S_i\leq4$.
Combining all three conditions for the three sectors of solutions, we notice that the ordered pair
$ (s_1,\overline{s}_1, s_2,\overline{s}_2, S_3)$ must form a pentagon. 
Again, we have checked with the methods of \cite{Costa:2011mg} and \cite{Elkhidir:2014woa} for the number of independent amplitudes and find agreement. 

\subsection{Two massive and one massless: equal masses}\label{sec:equalmass}

In this section, we keep $m_3=0$ but assume that the massive particles have the same mass: $m_1 = m_2 = m \neq 0$.
When the massive particles have the same mass, the spinor-helicity variables of the two massive spinors are linearly dependent, and the formalism presented so far must be modified. 
Due to the kinematic configuration, many of the two brackets ($2\times2$ tensors) of mass dimension one are only half-rank. 
Consequently, we can express them as a product of two two-component objects with only little group indices. 
This strategy has already been implemented in the six-dimensional spinor helicity formalism of \cite{Cheung:2009dc} (and the five-dimensional massless formalism of \cite{Zhao:2023}). 
However, before introducing these new variables, we take a more pedestrian approach and later show how the two approaches are equivalent. 

\paragraph{The pedestrian road.}
We start by introducing a reference momentum $q$ such that the massless polarization vector takes the form
\begin{align}
	\varepsilon^\mu_{3,i} = \frac{
		 \langle 3_I \vert q | \Gamma^\mu \vert 3_{J} \rangle
	}{
		p_3 \cdot q
	}
	\Gamma^{IJ}_i
	\,,
\end{align}
where $\Gamma^{IJ}_{\mu=i}$ is an $\SO(3)$ gamma matrix.
While we will not use the polarization vector as written above, we will use another related tensor generated by the reference momenta $q$. 
This tensor is a new independent object that can enter into the tensor building blocks. 

As usual, we enumerate the allowed tensor structures and form an ansatz for the possible three-point amplitudes with $(m_1,m_2,m_3) = (m,m,0)$.
There are no allowed mass dimension zero structures, and all of the mass dimension one and two building blocks in \eqref{eq:5dset0mm} carry over.
However, unlike in section \ref{sec:2M10}, there is a single mass dimension three tensor\footnote{We ignore the scalar denominator when counting the mass dimension of this tensor.} built from a reference momentum $q$ 
\begin{align}\label{eq:0mmeset2}
   \frac{ \langle \ul{3}|1|q|\ul{3}\rangle}{p_3\cdot q}.
\end{align}
where the denominator makes this structure invariant under the gauge transformation of $q$.
We will often gauge-fix $q$ such that $q \cdot p_3 = 1$ for simplicity.
This tensor satisfies the following relations:
\begin{equation}\label{eq:identity31q3}
    \begin{aligned}
        \langle \bul{1} \ul{3} \rangle \langle \bul{2}\ul{3}\rangle 
        = -\frac{1}{2} \langle\bul{12}\rangle \frac{\langle \ul{3}|1|q|\ul{3}\rangle}{p_3\cdot q}
        \,,
        &\qquad
        \lbrack \bul{1}\ul{3}\rangle \lbrack \bul{2}\ul{3}\rangle 
        = -\frac{1}{2} \lbrack \bul{12}\rbrack \frac{\langle \ul{3}|1|q|\ul{3}\rangle}{p_3\cdot q} 
        \,,
        \\
        \lbrack \bul{1}\ul{3}\rangle \langle \bul{1}\ul{3}\rangle 
        = \frac{i}{4m} \lbrack \bul{1}|3|\bul{1} \rangle \frac{\langle \ul{3}|1|q|\ul{3}\rangle}{p_3\cdot q} 
        \,,
        &\qquad
        \lbrack \bul{2}\ul{3}\rangle \langle \bul{2}\ul{3}\rangle 
        = -\frac{i}{4m} \lbrack \bul{2}|3|\bul{2} \rangle \frac{\langle \ul{3}|2|q|\ul{3}\rangle}{p_3\cdot q}
        \,,
    \end{aligned}
\end{equation}
that follow from generalized Schouten identities in five dimensions. 
One can see that (\ref{eq:identity31q3}) has divided the four brackets appearing on the LHS into two pairs
\begin{align}
    \{ \langle \bul{1}\ul{3}\rangle, \lbrack \bul{2}\ul3\rangle \} 
    \,,\quad 
    \{ \lbrack \bul1\ul3 \rangle, \langle \bul2\ul3 \rangle \},
\end{align}
where any two brackets from different pairs can be replaced by $\frac{\langle \ul3|1|q|\ul3\rangle}{p_3\cdot q}$. Therefore, we can write all amplitudes using  $\frac{\langle \ul3|1|q|\ul3\rangle}{p_3\cdot q}$ along with only one set of the two pairs.

One might be worried whether we have exhausted all the little group covariant structures. For example, the bracket $\langle \bul1 | q | \ul3 \rangle$ is definitely not covariant, but could any polynomial made up of this type of structure become covariant? To study this systematically, we have to define the $uw$-variables of \cite{Cheung:2009dc, Zhao:2023}.

\paragraph{$uw$-variables: ensuring spinor-space is spanned.}

Due to the kinematic configuration, the following $2\times2$ tensors of mass dimension one are only half-rank
\begin{equation}\label{eq:0mmhalflist}
    \begin{aligned}
        \langle \bul1 \ul3 \rangle \,,\quad \lbrack \bul1 \ul3 \rangle \,, \quad \langle \bul2 \ul3 \rangle \,, \quad \lbrack \bul2 \ul3 \rangle \,, \quad \langle \bul1 \bul2 \rangle \,, \quad \lbrack \bul{1 2} \rbrack \,.
    \end{aligned}
\end{equation}
On the other hand, the tensors $\langle \bul{1 2} \rbrack$ and $\lbrack \bul{1 2} \rangle$ have \emph{full} rank. 
Since the tensors listed in (\ref{eq:0mmhalflist}) all have rank one, we can express them as a product of two two-component objects with only little group indices. 
However, the same spinor helicity variables $|3\rangle$ appearing in different tensors might correspond to different two-component objects. 
For example, tensors $\langle \bul 1 \vert 3_I \rangle$ and $\langle \bul 2 \vert 3_I \rangle$ should be decomposed into different spinors with label $I$, since contracting the spinor helicity variable $|3_I\rangle$ in these two tensors is non-vanishing: $\langle \bul1 \vert 3^I \rangle \langle 3_I \vert \bul2 \rangle = 2im \langle \bul{12} \rangle$. Therefore, we find the following decomposition
\begin{equation}
    \begin{aligned}
        & \langle \mbf{1}_{\alpha} \vert 3_{I} \rangle 
        = \bs{u}_{1,\alpha} \bar{u}_{3,I} 
        \,, 
        && 
        \lbrack \mbf{1}_{\dot\alpha} \vert 3_{I} \rangle 
        = \bar{\bs{u}}_{1,\dot\alpha} u_{3,I} 
        \,, 
        \\
        & \lbrack \mbf{2}_{\dot\beta} \vert 3_{I} \rangle
        = \bs{u}_{2,\dot\beta} \bar{u}_{3,I} 
        \,, 
        && 
        \langle \mbf{2}_{\beta} \vert 3_{I} \rangle 
        = \bar{\bs{u}}_{2,\beta} u_{3,I} 
        \,, 
        \\
        & 
        \langle \mbf{1}_{\alpha} \vert \mbf{2}_{\beta}  \rangle
        = \bs{u}_{1,\alpha} \bar{\bs{u}}_{2,\beta} 
        \,, 
        && \lbrack \mbf{1}_{\dot\alpha} \vert \mbf{2}_{\dot\beta}  \rbrack
        = \bar{\bs{u}}_{1,\dot\alpha} \bs{u}_{2,\dot\beta} 
        \,.
    \end{aligned}
\end{equation}
Note that there is a notational discrepancy in defining $u$ variables for particles 1 and 2, where we use $u_1$ to denote Weyl spinors and $u_2$ to denote anti-Weyl spinors.

Next, we note that $u_3$ and $\bar{u}_3$ are linear independent
\begin{equation}
    u_{3,I} \bar{u}_{3,J} - u_{3,J} \bar{u}_{3,I} = -2m \varepsilon_{IJ} 
    = -2mi C_{IJ}
    \,,
\end{equation}
and, therefore, fully span the spinor space for particle 3. 
For particles-1 and -2, we define the ``pseudo-inverses'' of the $u$-variables: $w_1, w_2, \bar{w}_1, \bar{w}_2$. 
These pseudo-inverses are defined by the following equations
\begin{equation}
    \bs{u}_{1,\alpha} \bar{\bs{w}}_1^{\alpha} = 1 
    \,, \quad 
    \bar{\bs{u}}_{1,\dot\alpha} \bs{w}_1^{\dot\alpha} = 1 
    \,, \quad 
    \bs{u}_{2,\dot\beta} \bar{\bs{w}}_2^{\dot\beta} = 1 
    \,, \quad 
    \bar{\bs{u}}_{2,\beta} \bs{w}_2^{\beta} = 1 
    \,.    
\end{equation}
and satisfy the following completeness relations 
\begin{align}\begin{aligned}
    \bs{u}_{1,[\alpha} \bar{\bs{w}}_{1,\beta]} 
    &= -\varepsilon_{\alpha\beta}
    = - C_{\alpha\beta}
    = \bar{\bs{u}}_{2,[\alpha} \bs{w}_{2,\beta]}
    \,,
    \\
    \bar{\bs{u}}_{1,[\dot\alpha} \bs{w}_{1,\dot\beta]} 
    &= -\varepsilon_{\dot\alpha\dot\beta}
    = - C_{\dot\alpha\dot\beta}
    =\bs{u}_{2,[\dot\alpha} \bar{\bs{w}}_{2,\dot\beta]}
    \,.
\end{aligned}\end{align}
Unfortunately, there is a gauge redundancy in the definition of the $\bs{w}$-variables
\begin{equation} \label{eq:w-gauge}
    \bs{w}_i \to \bs{w}_i + b_i \bar{\bs{u}}_i \,, \quad \bar{\bs{w}}_i \to \bar{\bs{w}}_i + \bar{b}_i \bs{u}_i \,,
\end{equation}
where $b$'s and $\bar{b}$'s are pure numbers that satisfy $\sum_i \bar{b}_i =0=\sum_i b_i$. 
Since the amplitude must be invariant under these gauge transformations, we conclude that the only allowed tensors, including $w$- and $\bar{w}$-spinors, are
\begin{equation}
    \bs{u}_{1,\alpha} \bar{\bs{w}}_{2,\beta} 
        + \bar{\bs{w}}_{1,\alpha} \bs{u}_{2,\beta}
    \,, 
    \quad 
    \bar{\bs{u}}_{1,\dot\alpha} \bs{w}_{2,\dot\beta}
        + \bs{w}_{1,\dot{\alpha}} \bar{\bs{u}}_{2,\dot\beta}
    \,.
\end{equation}
Written in terms of the usual spinor helicity variables, 
\begin{equation}
        \langle \bul{1 2} \rbrack = \bul{u}_1 \bul{\bar{w}}_2 + \bul{\bar{w}}_1 \bul{u}_2 
        \,,\quad 
        \lbrack \bul{1 2} \rangle = \bul{\bar{u}}_1 \bul{w}_2 + \bul{w}_1 \bul{\bar{u}}_2 \,,
\end{equation}
they are the familiar mass dimension one two-brackets.
Here, we have introduced the underlined notation for the auxiliary $u$- and $w$- spinors. 
Explicitly, 
\begin{align}\begin{aligned}
    \bul{u}_1 = \bs{u}_{1,\alpha}\ z_1^\alpha
    \,,
    &\qquad
    \bul{\bar{u}}_1 = \bs{u}_{1,\dot\alpha}\ z_1^{\dot\alpha}
    \,,
    &\qquad
    \bul{\bar{w}}_1 = \bar{\bs{w}}_{1,\alpha}\ z_1^{\alpha}
    \,,
    &\qquad
    \bul{w}_1 = \bs{w}_{1,\dot\alpha}\ z_1^{\dot\alpha}
    \,,
    \\
    \bul{\bar{u}}_2 = \bs{u}_{2,\alpha}\ z_2^{\alpha}
    \,,
    &\qquad
    \bul{u}_2 = \bs{u}_{2,\dot\alpha}\ z_2^{\dot\alpha}
    &\qquad
    \bul{w}_2 = \bs{w}_{2,\alpha}\ z_2^{\alpha}
    \,,
    &\qquad
    \bul{\bar{w}}_2 = \bar{\bs{w}}_{2,\dot\alpha}\ z_2^{\dot\alpha}
    \,,
    \\
    &\qquad
    \ul{u}_3 = u_{3,I}\ z_3^I
    \,,
    &\qquad
    \ul{\bar{u}}_3 = u_{3,I}\ z_3^{I}
    \,.
    &
\end{aligned}\end{align}

To obtain the full set of little group covariant terms, we also have to note that all the $u$ and $w$ variables admit a global rescaling redundancy
\begin{equation}
\begin{aligned}
    &\bul{u}_i \to a \bul{u}_i \,, \quad \bul{\bar{u}}_i \to \frac{1}{a} \bul{\bar{u}}_i \,,\quad \bul{w}_i \to a \bul{w}_i \,,\quad \bul{\bar{w}}_i \to \frac{1}{a} \bul{\bar{w}}_i \, \quad \text{for}\quad i=1,2\,,
    \\&
    \ul{u}_3 \to a \ul{u}_3 \,, \quad \ul{\bar{u}}_3 \to \frac{1}{a} \ul{\bar{u}}_3\,.
\end{aligned}
\end{equation}
Thus, every little group covariant structure must include the same number of bar- and unbar-variables. 
This leaves us with the following seeds to build an ansatz for the amplitude
\begin{equation}
    \bul{u}_1 \bul{\bar{w}}_2 
        + \bul{\bar{w}}_1 \bul{u}_2 
    \,, \quad 
    \bul{\bar{u}}_1 \bul{w}_2 
        + \bul{w}_1 \bul{\bar{u}}_2 
    \,, \quad 
    \bul{u}_i \bul{\bar{u}}_j \,,
\end{equation}
where $i=j$ is allowed and we drop the boldface typesetting on the $u$'s when $i,j=3$. 
Therefore, we see that the space of five-dimensional three-point amplitudes with two equal masses is similar to the space of massless three-point amplitudes in six dimensions \cite{Cheung:2009dc}. 
The only difference is that the inverse of $u_3$ is fixed and has no gauge redundancy.

For completeness, we give a dictionary between $uw$-variables defined here and the usual spinor helicity variables in Table~\ref{tab:5dqumap0mm}. 
In particular, note that the $q$-dependent tensor introduced in the pedestrian approach is a gauge invariant combination of the $u$'s:  $\langle {}_I 3 | 1 | q | 3_{J} \rangle / (p_3 \cdot q) = u_{3,I} \bar{u}_{3,J} + \bar{u}_{3,I} u_{3,J}$.
Since the $u$- and $w$-spinors span the spinor space for all particles, we see from Table~\ref{tab:5dqumap0mm} that \eqref{eq:5dset0mm} and \eqref{eq:0mmeset2} exhausts all of the independent tensor structures.

\begin{table}
    \centering
    \begin{tabular}[!ht]{c|c}
        in spinor helicity variables & in $uw$-variables \\ \hline
        $\langle \bul{1} \ul{3} \rangle$ & $\bul{u}_1 \ul{\bar{u}}_{3}$ \\
        $\lbrack \bul{1} \ul{3} \rangle$ & $\bul{\bar u}_1 \ul{u}_{3}$ \\
        $\lbrack \bul{2} \ul{3} \rangle$ & $\bul{u}_2 \ul{\bar u}_{3}$ \\
        $\langle \bul{2} \ul{3} \rangle$ & $\bul{\bar u}_2 \ul{u}_{3}$ \\
        $\langle \bul{1} \bul{2} \rangle$ & $\bul{u}_1 \bul{\bar u}_2$ \\
        $\lbrack \bul{1} \bul{2} \rbrack$ & $\bul{\bar u}_1 \bul{u}_2$ \\
        $\langle \bul{1} \bul{2} \rbrack$ & $\bul{u}_1 \bul{\bar w}_2 + \bul{\bar w}_1 \bul{u}_2$ \\
        $\lbrack \bul{1} \bul{2} \rangle$ & $\bul{\bar u}_1 \bul{w}_2 + \bul{w}_1 \bul{\bar u}_2$ \\
        $\langle \bul{1} | 3 | \bul{1} \rbrack$ & $\bul{u}_1 \bul{\bar u}_1$ \\
        $\lbrack \bul{2} | 3 | \bul{2} \rangle$ & $\bul{u}_2 \bul{\bar u}_2$ \\
        $\langle \bul{1} | q | 3 | \bul{2} \rbrack$ & $\bul{u}_1 \bul{\bar w}_2$ \\
        $\lbrack \bul{2} | q | 3 | \bul{1} \rangle = - \langle \bul{1} | q | 3 | \bul{2} \rbrack $ & $\bul{\bar w}_1 \bul{u}_2$ \\
        \rule{0pt}{4ex} $\dfrac{ \langle \ul{3} | 1 | q | \ul{3} \rangle }{p_1 \cdot q}$ & $\ul{u}_{3} \ul{\bar{u}}_{3} + \ul{\bar{u}}_{3} \ul{u}_{3}$ \\
        \rule{0pt}{5ex} $\dfrac{ \langle \ul{3} | \hat{1} | q | \ul{3} \rangle }{p_1 \cdot q}$ & $\ul{u}_{3} \ul{\bar{u}}_{3}$ \\
        \rule{0pt}{5ex} $\dfrac{ \langle \ul{3} | \bar{1} | q | \ul{3} \rangle }{p_1 \cdot q}$ & $\ul{\bar{u}}_{3} \ul{u}_{3}$ \\
    \end{tabular}
    \caption{The map between spinor helicity variables and $u,w$ representation in two massive case with equal mass where $|\hat{1}| := | 1 \rangle \langle 1 | $ and $ |\bar{1}|:= | 1 \rbrack \lbrack 1 |$.
    Furthermore, note that the correspondence in the above table is up to constant multiplicitive factors.
    }
    \label{tab:5dqumap0mm}
\end{table}

The $uw$-variables also yield a better understanding of the identities (\ref{eq:identity31q3}). 
All the brackets appearing there are half-rank, and decomposing them into $u$ variables trivializes the identities. 
In other words, these identities are just collecting pairs of $(u_3,\bar{u}_3)$ together to form $\frac{\langle 3 | 1 | q | 3 \rangle}{p_3 \cdot q}$. 

\paragraph{The amplitudes.}
Now that we are confident that we are not missing any tensors, we can build an ansatz for the amplitudes using either the pedestrian or $uw$ variables. 
We will proceed with the construction in $uw$-variables. 
As in section \ref{sec:2M10}, the spin of  particle-1 and -2 are labeled by $(s_1, \bar{s}_1)$ and $(s_2, \bar{s}_2)$ while $S_3$ labels the spin of particle-3. 
Also we define $S_1 = s_1 + \bar{s}_1$ and $S_2 = s_2 + \bar{s}_2$. 
Then, little group covariance implies that
\begin{equation} \label{eq:m00ansatz}
        A(\mbf{1}^{s_1+\bar{s}_1}, \mbf{2}^{s_2 + \bar{s}_2}, 3^{S_3}) = \left( \bul{u}_1 \bul{\bar{w}}_2 + \bul{\bar{w}}_1 \bul{u}_2 \right)^{v_0} \left( \bul{\bar{u}}_1 \bul{w}_2 + \bul{w}_1 \bul{\bar{u}}_2 \right)^{\bar{v}_0} \bul{u}_1^{v_1} \bul{\bar{u}}_1^{\bar{v}_1} \bul{u}_{2}^{v_2} \bul{\bar{u}}_2^{\bar{v}_2} \ul{u}_{3}^{v_3} \ul{\bar{u}}_{3}^{\bar{v}_3} 
\end{equation}
with 
\begin{equation}
    \begin{aligned}
        v_0 + v_1 = 2s_1 
        \,,\qquad
        & \bar{v}_0 + \bar{v}_1 = 2\bar{s}_1 
        \,,\qquad
        && v_0 + v_2 = 2\bar{s}_2 
        \,,
        \\
        \bar{v}_0 + \bar{v}_2 = 2s_2 
        \,,\qquad
        & v_3 + \bar{v}_3 = 2S_3 
        \,, \qquad
        && v_1 + v_2 + v_3 = \bar{v}_1 + \bar{v}_2 + \bar{v}_3 \,.
    \end{aligned} 
\end{equation}
Since there are eight indeterminates but only six equations, the solution space of these equations is two-dimensional. 
In the following, we choose $v_0$ and $\bar{v}_0$ to be free and fix all other variables.

Imposing all $v$'s and $\bar{v}$'s to be non-negative puts constraints on both the spins and the tuple $(v_0,\bar{v}_0)$. 
On the one hand, the constraints on spin configuration are

\begin{equation}
    \left\{
    \begin{aligned}
        S_3 + \bar{s}_1 + s_2 \geqslant |s_1 - \bar{s}_2| \,,\\
        S_3 + s_1 + \bar{s}_2 \geqslant |\bar{s}_1 - s_2| \,,
    \end{aligned} \right.
\end{equation}
and imply that any number among $s_1, \bar{s}_1, s_2, \bar{s}_2$ cannot be larger than the sum of the rest plus $S_3$. Compared with the unequal mass case, these spin configurations can break the pentagon rule since the relation $S_3 \leqslant s_1+\bar{s}_1+s_2+\bar{s}_2$ is not required.
\begin{figure}
    \centering
    \begin{tikzpicture}[baseline={([yshift=-2ex]current bounding box.center)}]
        \draw[thick,->] (0,0) -- (0,3) node[left,Blue]{$\min(2\bar{s}_1,2s_2)$} -- (0,4) node[above]{$\bar{v}_0$};
        \draw[thick,->] (0,0) -- (3,0) node[above right,Blue]{$\min(2s_1,2\bar{s}_2)$} -- (5.5,0) node[right]{$v_0$};
        \draw[ultra thick,Blue] (3,0) -- (3,3) -- (0,3);
        \draw[ultra thick,dashed,Maroon] (-1.5,0.5) -- (0,2) (1,3) -- (2,4);
        \draw[ultra thick,dashed,Maroon] (0,-1) -- (1,0) (3,2) -- (4,3);

        \draw[line width=0,white,fill=gray!20] (1,3) -- (3,3) -- (3,2) -- (1,0) -- (0,0) -- (0,2) -- cycle;
        \draw[ultra thick,violet] (1,3) -- (3,3) -- (3,2) (0,2) -- (0,0) -- (1,0);
        \draw[ultra thick,violet,dashed] (1,3) -- (0,2) (3,2) -- (1,0);

        \foreach \x in {0,0.5,1,1.5,2,2.5,3}
        {
            \foreach \y in {0,0.5,1,1.5,2,2.5,3}
                \draw[ultra thick, fill=black!60] (\x,\y) circle (.8pt);
        };

        \draw[ultra thick,fill=Maroon,Maroon] (0,2) node[left]{$S_3-s_1+\bar{s}_1+s_2-\bar{s}_2 \ $} circle (1.5pt);
        \draw[ultra thick,fill=Maroon,Maroon] (1,0) node[below right=1pt]{$S_3+s_1-\bar{s}_1-s_2+\bar{s}_2$} circle (1.5pt);
    \end{tikzpicture}
    \caption{The allowed region of integer solutions for the exponents $v_0$ and $\bar{v}_0$ in the ansatz for the $(m_1,m_2,m_3)=(m,m,0)$ amplitudes in equation \eqref{eq:m00ansatz}.}
    \label{fig:0mmeregion}
\end{figure}
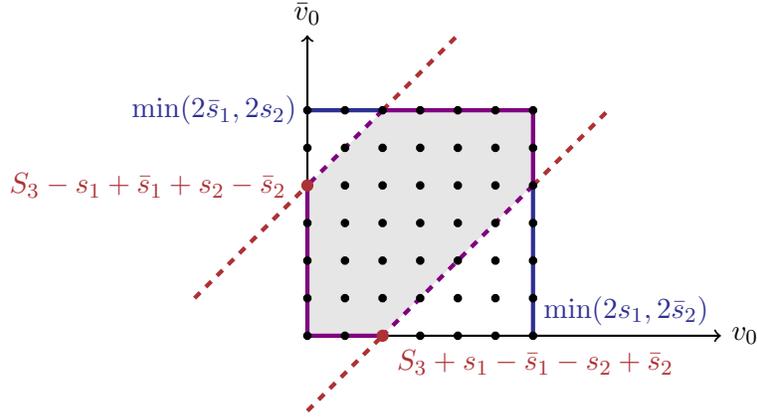

On the other hand, from the constraints on the tuple $(v_0,\bar{v}_0)$ we can obtain the number of independent couplings. 
As depicted in figure \ref{fig:0mmeregion}, the allowed values for $(v_0,\bar{v}_0)$ are the integer points in the shaded area (including the boundary). 
Note that the graph above just corresponds to one type of solution, in the sense that the relation between the two lines and the square is more general. For example, they can be either below or above the diagonal of the square or totally not intersecting with the square in some spin configurations.

\paragraph{Example ($\boldsymbol{WWA}$-scattering):} 
In our formalism, the amplitude for the scattering of two $W$-bosons (massive spin-one particles) with a massless spin-one particle ($A$) is 
\begin{align}
   A(W,W,A)=
   A(\mbf{1}^{\frac{1}{2}+\bar{\frac{1}{2}}},\mbf{2}^{\frac{1}{2}+\bar{\frac{1}{2}}},3^{1})=
   \frac{\langle \ul{3}|1|q|\ul{3}\rangle}{p_3\cdot q}\langle \bul{12}\rbrack \lbrack \bul{12} \rangle.
\end{align}
Using the identities \eqref{eq:identity31q3}, one can show that this is equivalent to the expression in \cite{Chiodaroli:2022ssi}.

\paragraph{Example ($\boldsymbol{\chi\bar{\chi}A}$- and $\boldsymbol{\chi\chi A}$-scattering):}
Here, we consider the scattering between massive spin-$\frac12$ particles ($\chi$ and $\bar{\chi}$) and a massless vector boson ($A$).
Depending on the chirality of the two fermions, there are two configurations. 

The first corresponds to a chirality-preserving interaction
\begin{align}
    &\begin{aligned}
        A(\chi,\bar{\chi},A)
        = A(\mbf{1}^{\frac{1}{2}},\mbf{2}^{\bar{\frac{1}{2}}},3^{1})
        &= g_{0,0} \bul{u}_1 \bul{u}_{2} \ul{\bar{u}}_{3}^{2} + g_{1,0} \left( \bul{u}_1 \bul{\bar{w}}_2 + \bul{\bar{w}}_1 \bul{u}_2 \right) \ul{u}_{3} \ul{\bar{u}}_3 \\
        &\sim g_{0,0} \langle \bul{1} \ul{3} \rangle \lbrack \bul{2} \ul{3} \rangle + g_{1,0} \langle \bul{1} \bul{2} \rbrack \langle \bul{3} | 1 | q | \bul{3} \rangle \\
        &\sim g_{0,0} \langle \bul{1} \ul{3} \rangle \lbrack \bul{2} \ul{3} \rangle + g_{1,0} \left( \langle \bul{1} \ul{3} \rangle \lbrack \bul{2} | q | \ul{3} \rangle - \lbrack \bul{2} \ul{3} \rangle \langle \bul{1} | q | \ul{3} \rangle \right) 
        \,.
    \end{aligned}
\end{align}
Here, we have absorbed any proportionality constants converting from the 
$uw$ variables to the normal spinor helicity variables into the couplings $g_{0,0}$ and $g_{1,0}$. 
In fact, the term proportional to $g_{1,0}$ corresponds to the standard interaction of two spinors with massless gauge bosons and is exactly the amplitude found in~\cite{Chiodaroli:2022ssi}. 
On the other hand, the term proportional to $g_{0,0}$ corresponds to some higher derivative deformations in Lagrangian. 

The second configuration 
\begin{align}
    \begin{aligned}\label{eq:chichia}
        A(\chi,\chi,A)
        =  A(\mbf{1}^{\frac{1}{2}},\mbf{2}^{\frac{1}{2}},3^{1})
        &= g'_{0,0} \bul{u}_1 \bul{\bar u}_{2} \ul{u}_{3} \ul{\bar{u}}_3 
        \sim g'_{0,0} \langle \bul{1} \bul{2} \rangle \langle \ul{3} | 1 | q | \ul{3} \rangle 
        \sim g'_{0,0} \langle \bul{1} \ul{3} \rangle \langle \bul{2} \ul{3} \rangle 
        \,
    \end{aligned}
\end{align}
does not preserve chirality. 
Moreover, no $g'_{1,0}$ term exists because gauge theory interactions always preserve chirality. 
On the other hand, the $g'_{0,0}$ term in this chirality-breaking configuration implies that there could be higher-derivative interactions that do not preserve chirality.

\paragraph{Arbitrary $(\bul{1}^{S},\bul{2}^{S},\ul{3}^{h})$-scattering.}
To study black hole scattering, we need to understand the configuration $s_1 = \bar{s}_1 = s_{2} = \bar{s}_2 = \frac{1}{2}S$, where we denote the spin of particle-1 and particle-2 by $S$. 
We also set $S_3 = h$ as the helicity of particle-3. 
According to figure \ref{fig:0mmeregion}, the independent amplitudes are given by the intersection of a square with edge length $S$ and the region between the two oblique lines $v_0 - \bar{v}_0 = h$ and $v_0 - \bar{v}_0 = -h$. 
When $S \leqslant h$, the full square is between the two oblique lines, so the allowed region for $(v_0,\bar{v}_0)$ is the full square, and there are $(S+1)^2$ independent couplings
\begin{equation} \label{eq:SSh1}
    \begin{aligned}
        &A(\bul{1}^{S},\bul{2}^{S},\ul{3}^{h}) 
        = \sum_{v_0, \bar{v}_0 = 0}^{S} 
        g_{v_{0},\bar{v}_0} 
        \left( \bul{u}_1 \bul{\bar{w}}_2 {+} \bul{\bar{w}}_1 \bul{u}_2 \right)^{v_0} 
        \left( \bul{\bar{u}}_1 \bul{w}_2 {+} \bul{w}_1 \bul{\bar{u}}_2 \right)^{\bar{v}_0} 
        \\&\hspace{10em}\times
        \bul{u}_1^{S-v_0} 
        \bul{\bar{u}}_1^{S-\bar{v}_0} 
        \bul{u}_{2}^{S-v_0} 
        \bul{\bar{u}}_2^{S-\bar{v}_0} 
        \ul{u}_{3}^{h+v_0-\bar{v}_0} 
        \ul{\bar{u}}_{3}^{h-v_0+\bar{v}_0} 
        \\
        &= \sum_{\substack{v_0< \bar{v}_0 = 0 \\ v_0 < \bar{v}_0}}^{S} g_{v_0,\bar{v}_0} \langle \bul{1} \bul{2} \rbrack^{v_0} \lbrack \bul{1} \bul{2} \rangle^{\bar{v}_0} \langle \bul{1} \bul{2} \rangle^{S-\bar{v}_0} \lbrack \bul{1} \bul{2} \rbrack^{S-\bar{v}_0} \langle \bul{1} \ul{3} \rangle^{\bar{v}_0 - v_0} \lbrack \bul{2} \ul{3} \rangle^{\bar{v}_0 - v_0} \langle \ul{3} | 1 | q | \ul{3} \rangle^{h+v_0-\bar{v}_0} \\
        &\quad
        +\sum_{\substack{v_0, \bar{v}_0 = 0 \\ v_0 \geqslant \bar{v}_0}}^{S} g_{v_0,\bar{v}_0} \langle \bul{1} \bul{2} \rbrack^{v_0} \lbrack \bul{1} \bul{2} \rangle^{\bar{v}_0} \langle \bul{1} \bul{2} \rangle^{S-v_0} \lbrack \bul{1} \bul{2} \rbrack^{S-v_0} \lbrack \bul{1} \ul{3} \rangle^{v_0 - \bar{v}_0} \langle \bul{2} \ul{3} \rangle^{v_0 - \bar{v}_0} \langle \ul{3} | 1 | q | \ul{3} \rangle^{h-v_0+\bar{v}_0} 
        \,.
        \\
    \end{aligned}
\end{equation}
When $S>h$, the available region then becomes a hexagon and there are $(S+1)^2 - (S-h)^2 - (S-h)$ independent couplings
\begin{equation}\label{eq:SSh2}
    \begin{aligned}
        &A(\bul{1}^{S},\bul{2}^{S},\ul{3}^{h})
        = \sum_{\substack{v_0, \bar{v}_0 = 0 \\ |v_0 - \bar{v}_0| \leqslant h}}^{S} g_{v_{0},\bar{v}_0} \left( \bul{u}_1 \bul{\bar{w}}_2 {+} \bul{\bar{w}}_1 \bul{u}_2 \right)^{v_0} \left( \bul{\bar{u}}_1 \bul{w}_2 {+} \bul{w}_1 \bul{\bar{u}}_2 \right)^{\bar{v}_0} 
        \\&\hspace{10em}
        \times \bul{u}_1^{S-v_0} \bul{\bar{u}}_1^{S-\bar{v}_0} \bul{u}_{2}^{S-v_0} \bul{\bar{u}}_2^{S-\bar{v}_0} \ul{u}_{3}^{h+v_0-\bar{v}_0} \ul{\bar{u}}_{3}^{h-v_0+\bar{v}_0} \\
        &= \sum_{\substack{v_0< \bar{v}_0 = 0 \\ v_0 < \bar{v}_0 \leqslant v_0+h}}^{S} g_{v_0,\bar{v}_0} \langle \bul{1} \bul{2} \rbrack^{v_0} \lbrack \bul{1} \bul{2} \rangle^{\bar{v}_0} \langle \bul{1} \bul{2} \rangle^{S-\bar{v}_0} \lbrack \bul{1} \bul{2} \rbrack^{S-\bar{v}_0} \langle \bul{1} \ul{3} \rangle^{\bar{v}_0 - v_0} \lbrack \bul{2} \ul{3} \rangle^{\bar{v}_0 - v_0} \langle \ul{3} | 1 | q | \ul{3} \rangle^{h+v_0-\bar{v}_0} \\
        &\quad 
        {+}\sum_{\substack{v_0, \bar{v}_0 = 0 \\ \bar{v}_0 \leqslant v_0 \leqslant \bar{v}_0+h}}^{S} g_{v_0,\bar{v}_0} \langle \bul{1} \bul{2} \rbrack^{v_0} \lbrack \bul{1} \bul{2} \rangle^{\bar{v}_0} \langle \bul{1} \bul{2} \rangle^{S-v_0} \lbrack \bul{1} \bul{2} \rbrack^{S-v_0} \lbrack \bul{1} \ul{3} \rangle^{v_0 - \bar{v}_0} \langle \bul{2} \ul{3} \rangle^{v_0 - \bar{v}_0} \langle \ul{3} | 1 | q | \ul{3} \rangle^{h-v_0+\bar{v}_0} 
        \,.
        \\
    \end{aligned}
\end{equation}

\subsection{One massive and two massless \label{sec:1M20}}

In this section, we set $m_1=m_2=0$, and $m_3 \neq 0$ (note that particle-3 is massive now!). 
Since all brackets are full rank, we do not need to introduce $uw$-variables and can proceed as in sections \ref{sec:genM} and \ref{sec:2M10}. 
For this kinematic configuration, the space of amplitudes splits into two sectors
\begin{align} \label{eq:A3-1M20}
    A(\ul{1}^{S_1}, \ul{2}^{S_2}, \bul{3}^{s_3 + \bar{s}_3}) &=
    \begin{cases}
        \begin{aligned}
            \sum_{v \in R_1} g_{v} \bigg[&
            \langle \ul{1} \bul{3} \rangle^{v}
            \langle \ul{2} \bul{3} \rangle^{s_3-\bar{s}_3+S_1+S_2-v}
            \langle \ul{1} \bul{3} ]^{2S_1-v,} 
            \\&\quad\times 
            \langle \ul{2} \bul{3} ]^{\bar{s}_3-s_3-S_1+S_2+v} 
            \bigg] \langle \bul{3} |1| \bul{3} ]^{S_3-S_1-S_2}
        \end{aligned}
        & \text{ if } S_3 \geq S_1+S_2
        \,, 
        \\
        \begin{aligned}
            \sum_{v\in R_2} g_{v} \bigg[&
            \langle \ul{1} \bul{3} \rangle^{v}
            \langle \ul{2} \bul{3} \rangle^{2s_3-v} 
            \langle \ul{1} \bul{3} ]^{s_3 + \bar{s}_3+S_1-S_2-v} 
            \\&\quad\times
            \langle \ul{2} \bul{3} ]^{\bar{s}_3-s_3-S_1+S_2+v} 
            \bigg] \la\ul{12}\ra^{S_1+S_2 - S_3}
        \end{aligned}
         & \text{ if } S_1+S_2 \geq S_3
         \,, 
    \end{cases}
\end{align}
where  
\begin{align}
    R_1 &= \left[ 
        \max\left(0, s_3{-}\bar{s}_3{+}S_1{-}S_2\right),
        \min\left(2S_1,s_3{-}\bar{s}_3{+}S_1{+}S_2\right)
        \right] \cap \mathbb{Z}
    \\
    R_2 &= \left[
        \max\left(0, s_3{-}\bar{s}_3{+}S_1{-}S_2\right),
        \min\left(s_3,s_3{+}\bar{s}_3{+}S_1{-}S_2\right)
        \right] \cap \mathbb{Z}
\end{align}
are the lattice points inside a line segment. 
Note that for this formula to make sense, the $\{S_1,S_2,s_3,\bar{s}_3\}$ must satisfy additional constraints
\begin{equation} \label{eq:Sconstraints-1M20}
    \begin{aligned}
        & S_1+S_2+S_3 \in \mathbb{N}_{\geq0} \,, \qquad
        && S_1 + S_2 + s_3 \geq \bar{s}_3 \,, \qquad
        & S_1 + S_2 + \bar{s}_3 \geq s_3 \,, 
        \\
        & S_1 + s_3 + \bar{s}_3 \geq S_2 \,, \qquad
        && S_2 + s_3 + \bar{s}_3 \geq S_1 . 
        &
    \end{aligned}
\end{equation}
These constraints mean that there are an even number of spinor brackets and that the $\{S_1,S_2,s_3,\bar{s}_3\}$ form a (possibly degenerate) quadrilateral on the plane. 
Thus, for any set $\{S_1,S_2,s_3,\bar{s}_3\}$ that satisfies \eqref{eq:Sconstraints-1M20} the number of independent couplings is simply the size of the set $|R_1|$ when $S_3\geq S_1+S_2$ and $|R_2|$ when $S_3 < S_1+S_2$.
We have performed an explicit numerical check that all of the above component amplitudes are linearly independent for $S_i\leq4$.

\paragraph{Example ($\boldsymbol{\psi \psi W}$-scattering):} 
Consider the scattering of two massless fermions $\psi$ ($S_1=S_2=\frac12$) with a massive $W$-boson ($S_3 = \frac12 + \bar{\frac12}$). Then, equation \eqref{eq:A3-1M20} becomes
\begin{align}
    A(\psi, \psi, W) = 
    A(\ul{1}^{\frac12}, \ul{2}^{\frac12}, \bul{3}^{\frac12 + \bar{\frac12}}) &= 
    g_1\
    \langle \ul{1}\bul{3} \rangle \langle \ul{2}\bul{3} ]
    + g_2\ 
    \langle \ul{2}\bul{3} \rangle \langle \ul{1}\bul{3} ]
    .
\end{align}
It is then easy to verify that the above two-component amplitudes are linearly independent. 

\paragraph{Example (higher spin bosons):} 
Consider the scattering of higher spin bosons with $S_1=S_2=4$ and $S_3=4 = 2+\bar{2}$. Then, \eqref{eq:A3-1M20} becomes
\begin{align}
\begin{aligned}
    A(\ul{1}^{4}, \ul{2}^{4}, \bul{3}^{2 + \bar{2}}) &= 
    g_1\
    \la\ul{12}\ra^4
    \la\ul{1}\bul{3}\ra^4
    \la\ul{2}\bul{3}]^4
    + g_2\
    \la\ul{12}\ra^4
    \la\ul{1}\bul{3}\ra^3
    \la\ul{1}\bul{3}]
    \la\ul{2}\bul{3}]^3
    \la\ul{2}\bul{3}\ra
    \\&\qquad
    + g_3\
    \la\ul{12}\ra^4
    \la\ul{1}\bul{3}\ra^2
    \la\ul{1}\bul{3}]^2
    \la\ul{2}\bul{3}]^2
    \la\ul{2}\bul{3}\ra^2
    + g_4\
   \la\ul{12}\ra^4
    \la\ul{1}\bul{3}\ra
    \la\ul{1}\bul{3}]^3
    \la\ul{2}\bul{3}]
    \la\ul{2}\bul{3}\ra^3
    \\&\qquad
    + g_5\
    \la\ul{12}\ra^4
    \la\ul{1}\bul{3}]^4
    \la\ul{2}\bul{3}\ra^4
    \,.
\end{aligned}
\end{align}
One can also show that these five component amplitudes are linearly independent.

\subsection{Three massless \label{sec:3massless}}

In this section, we set all masses to zero: $m_1=m_2=m_3=0$. 
Then, all particles are associated to an $\SO(3)$-bracket $\vert\ul{i}\ra$. 
As usual, the first step is to write down the allowed amplitudes, we enumerate all possible tensor structures. 
Because this kinematic configuration is degenerate, we make some of the steps explicit. 
The full set of independent building blocks for all-massless amplitudes can be found in table \ref{tab:5dqumap} in both spinor helicity and $uw$-variables.

At mass dimension one, we have the following basic tensors two-brackets $\la\ul{ij}\ra$. 
However, since the fully massless configuration is kinematically degenerate, we need to introduce new mass dimension one tensors called $x$-factors that parameterize the ratio of two collinear spinor brackets. 
For example, in the  all-massless configuration, the spinors $|1|3\rangle$ and $|3\rangle$ are parallel
\begin{equation}\label{eq:5masslesslist0}
    \langle 3_{I} | 1 | 3_{J} \rangle \sim \varepsilon_{IJ} p_1 \cdot p_3 = 0 \,.
\end{equation}
Therefore, we define their ``ratio'' as a new kinematic variable
\begin{equation}
    (x_i)_{IJ} 
    := \left( \langle i \, q_{i,I} \rangle^{-1} \right)^{K} \langle q_{i,K} | p_{i+1} | i_J \rangle 
    = \frac{ \langle i_I | q_i | p_{i+1} | i_{J} \rangle}{p_i \cdot q_i} \propto u_{i,I} u_{i,J}\,.
\end{equation}
where $q_i$ is an arbitrary reference vector. In index-free notation this is
\begin{equation}\label{eq:5masslesslist2}
    x_i = \langle \ul{i} | q_{i} | p_{i+1} | \ul{i} \rangle \propto \ul{u_i} \ul{u_i}. 
\end{equation}
This $x$-factor is simply a five-dimensional generalization of the four-dimensional $x$-factor of \cite{Arkani-Hamed:2017jhn,Zhao:2023}. The fact that it degenerates and is not an additional independent tensor is a reflection of the fact that minimal coupling is different in five dimensions than in four \cite{Chung:2019yfs, Zhao:2023}.

At mass dimension two, there are no non-vanishing tensors 
that do not contain the reference momenta $q_i$ since 
\begin{equation}
    \begin{aligned}
        \langle \ul{3} | 1 | \ul{3} \rangle = 0 
        \quad\text{and}\quad
        \langle \ul{3} | 1 | \ul{2} \rangle = - \langle \ul{3} | 2 | \ul{2} \rangle - \langle \ul{3} | 1 | \ul{2} \rangle = 0
        \,.
    \end{aligned}
\end{equation}
At mass dimension three, we find one gauge invariant (QED) term
\begin{equation}\label{eq:5masslesslist3}
    \langle \ul{1} \ul{3} \rangle \langle \ul{2} | q_3 | \ul{3} \rangle - \langle \ul{1} | q_3 | \ul{3} \rangle \langle \ul{2} \ul{3} \rangle \,,
\end{equation}
that cannot be represented in terms of $x$-factors and two-brackets $\la\ul{ij}\ra$. 

Without performing an analysis counting the number of independent amplitudes, we cannot be sure that we have exhausted all of the independent tensor structures. 
However, as shown in \cite{Zhao:2023}, the $uw$-variables span all of spinor space. 
Therefore, by providing a map from the spinor helicity variables to the $uw$ variables in table~\ref{tab:5dqumap}, we are sure we have not missed any tensor structures. 
In the end, the full list of tensor structures includes the two-brackets $\la\ul{ij}\ra$,  (\ref{eq:5masslesslist3}) along with a Yang-Mills term \eqref{eq:YM}.

\begin{table}[]
    \centering
    \begin{tabular}{c|c}
        $q$-representation & $uw$-variables \\ \hline
        $\langle \ul{1} \ul{3} \rangle$ & $\ul{u}_{1} \ul{u}_{3}$ \\
        $\langle \ul{3} \ul{1} \rangle$ & $- \ul{u}_{3} \ul{u}_{1}$ \\
        $\langle \ul{1} | q_1 | 2 | \ul{1} \rangle$ & $\ul{u}_{1} \ul{u}_{1}$ \\
        $\frac{ \langle \ul{1} | q_1 - (p_3 \cdot q_{1}) q_{3} | \ul{3} \rangle}{1 - (p_3 \cdot q_1)(p_1 \cdot q_3)}$ & $\ul{u}_1 \ul{w}_3$ \\
        $ \langle \ul{1} \ul{3} \rangle \langle \ul{2} | q_3 | \ul{3} \rangle - \langle \ul{2} \ul{3} \rangle \langle \ul{1} | q_3 | \ul{3} \rangle $ & $( \ul{u}_{1} \ul{u}_{2} \ul{w}_{3} + \ul{u}_{1} \ul{w}_{2} \ul{u}_{3} + \ul{w}_{1} \ul{u}_{2} \ul{u}_{3} ) \ul{u}_{3}$
    \end{tabular}
    \caption{The map between $q$ representation and $u,w$ representation.}
    \label{tab:5dqumap}
\end{table}

\paragraph{Example ($\boldsymbol{AAA}$-scattering in Yang-Mills):}
For massless matter, our spinor-helicity variables are equivalent to those of  \cite{Chiodaroli:2022ssi}. The Yang-Mills 3-gluon amplitude is 
\begin{equation} \label{eq:YM}
    A(A,A,A)
    = A(\ul{1}^1, \ul{2}^1, \ul{3}^1)  
    = \langle\ul{2q_3}\rangle 
    \langle\ul{3q_2}\rangle 
    \langle\ul{1}\vert k_2 \vert \ul{q_1} \rangle
    + \text{cyclic}
\end{equation}
This can be written as $(\ul{u}_1 \ul{u}_2 \ul{w}_3 + \ul{u}_1 \ul{w}_2 \ul{u}_3 + \ul{w}_1 \ul{u}_2 \ul{u}_3)^2$ in $uw$-variables \cite{Zhao:2023, Chiodaroli:2022ssi}.

\paragraph{Example ($\boldsymbol{\psi\psi A}$-scattering in massless QED):}
The coupling of massless fermions ($\psi$) to a massless vector boson ($A$) is 
\begin{equation}
    A(\psi,\psi,A)=
    A(\ul{1}^{\frac12}, \ul{2}^{\frac12}, \ul{3}^1) = \langle \ul{1} \ul{3} \rangle \langle \ul{2} | q_3 | \ul{3} \rangle - \langle \ul{1} | q_3 | \ul{3} \rangle \langle \ul{2} \ul{3} \rangle \,.
\end{equation}
This is (\ref{eq:5masslesslist3}) and can be written as $(\ul{u}_1 \ul{u}_2 \ul{w}_3 + \ul{u}_1 \ul{w}_2 \ul{u}_3 + \ul{w}_1 \ul{u}_2 \ul{u}_3) \ul{u}_3$ in the massless $uw$-variables \cite{Zhao:2023, Chiodaroli:2022ssi}.

\section{High energy limit \label{sec:HElim}}

In this section, we provide a prescription for taking the high energy limit of the massive spinor helicity variables (section \ref{sec:SH_HE}) and massive amplitudes (section \ref{sec:Amp_HE}). 
In particular, intricate cancellations are often required to arrive at gauge invariant amplitudes in the high energy limit (see \eqref{eq:Wphiphi_HE} and \eqref{eq:g-}). 
Lastly, at the end of section \ref{sec:Amp_HE}, we show how spin and mass dimensions of massive amplitudes are not preserved when taking the high energy limit and propose a way in which the massive amplitudes can be categorized by their high energy limit. 

\subsection{High energy limit of the spinor helicity variables \label{sec:SH_HE}}

Since any massive momentum $p^\mu$ ($p^2=-m^2$) can be written as the sum of two null momenta $k^\mu,q^\mu$
\begin{equation}
    p^{\mu} = k^{\mu} + m^2 q^{\mu} \,,
    \qquad
    k\cdot q = -\frac12\,,
\end{equation}
one can decompose the massive spinor helicity variables into massless $k$ and $q$ spinor helicity variables
\begin{align}\label{eq:p2kq}
    &\vert p_\alpha\ra 
    = (\vert k_I\ra + m \vert q_I\ra)
    U^I_{\enspace\alpha}\,,
    \qquad
    \vert p_{\dot{\alpha}}]
    = (\vert k_I\ra - m \vert q_I\ra)
    U^I_{\enspace\dot{\alpha}}\,,
\end{align}
where 
\begin{align}
    \label{eq:qkrelation}
    \langle k_{I} | q_{J} \rangle 
    =\begin{pmatrix}
        0 & 1
        \\
        -1 & 0
    \end{pmatrix} 
    = \varepsilon_{IJ}\,,
    \quad \text{and} \quad 
    U^I_{\enspace\alpha} 
    = U^I_{\enspace\dot{\alpha}} 
    =\frac{1}{\sqrt{2}}
    \begin{pmatrix}
        i & 0 
        \\
        0 & 1
    \end{pmatrix}
    \,.
\end{align}
Here, $U^I_\alpha$ and $U^I_{\dot{\alpha}}$ are a little group conversion matrices that transform the $\SO(4)$ $C$-matrix into the $\SO(3)$ $C$-matrix
\begin{align}
    U^I_{\enspace\dot{\alpha}} \
    U^J_{\enspace\dot{\beta}} \
    C^{\dot{\beta}\dot{\alpha}} 
    &= U^I_{\enspace\dot{\alpha}} \
    U^J_{\enspace\dot{\beta}} \
    \epsilon^{\dot{\beta}\dot{\alpha}} 
    = - \frac{i}{2} \epsilon^{IJ}
    = - \frac{1}{2} C^{IJ}
    \,,
    \\
    U^I_{\enspace\alpha} \
    U^J_{\enspace\beta} \
    C^{\beta\alpha} 
    &= U^I_{\enspace\alpha} \
    U^J_{\enspace\beta} \
    \epsilon^{\beta\alpha}
    = - \frac{i}{2} \epsilon^{IJ}
    = -\frac{1}{2} C^{IJ}
    \,.
\end{align}
While the $U$-matrices are strictly needed, we will often drop them because they can be easily restored. 

To find an explicit parameterization for the $k$ and $q$ spinors, we fix
\begin{equation}\label{eq:highepara}
    p^{\mu} = (E, \sqrt{E^2 - m^2} ,0,0,0) \,.
\end{equation}
Then, a convenient choice for $k$ and $q$ is 
\begin{equation}\label{eq:higheparakq}
    \begin{aligned}
        & k^{\mu} = \frac{1}{2} \left( E + \sqrt{E^2 - m^2}, E + \sqrt{E^2 - m^2}, 0,0,0 \right) \,, \\
        & q^{\mu} = \frac{1}{2m^2} \left( E - \sqrt{E^2 - m^2}, \sqrt{E^2 - m^2} - E, 0,0,0 \right) \,. 
    \end{aligned}
\end{equation}
In the high energy limit $E \gg m$, $p^\mu$ becomes approximately null and $k^\mu \gg q^\mu$. 
The $k$-spinors are given by substituting $k$ into (\ref{eq:mless-spinors}). 
On the other hand, the representation (\ref{eq:mless-spinors}) for $q$ is not allowed since $q_0+q_1=0$ and  we use an $\SO(3)$ rotation of \eqref{eq:mless-spinors} instead
\begin{equation} \label{eq:HEqspinor}
    \vSpinor_{A}|q_{I}\rangle = i \begin{pmatrix}
        0 & \frac{\sqrt{E^2-m^2}-E}{m^2} \\
        0 & 0 \\
        0 & 0 \\
        1 & 0 \\
    \end{pmatrix}
    \,.
\end{equation}

\subsection{High energy limit of amplitudes \label{sec:Amp_HE}}

In this section, we consider several examples to illustrate how to take the high energy limit of massive amplitudes. In particular, we demonstrate that intricate cancellations are often needed to produce gauge invariant massless amplitudes. 

\paragraph{Example (high energy limit of $\boldsymbol{\chi\chi\phi}$-scattering):}
Consider $\chi\chi\phi$-scattering where $\chi$ is a massive fermion ($s_1=\frac{1}{2}$ and $\bar{s}_1=0$) and $\phi$ is a massless scalar. 
According to Section~\ref{sec:equalmass}, there is only one independent term whose high energy limit is straightforward to take
\begin{equation}
    A(\chi,\chi,\phi) = 
    A(\mbf{1}^{\frac{1}{2}}, \mbf{2}^{\frac{1}{2}}, {3}^0) = 
    \langle \mathbf{1}_{\alpha} \vert \mathbf{2}_{\beta} \rangle \overset{m \ll E}{\longrightarrow} \langle k_{1,I} \vert k_{2,J} \rangle 
    U^I_{\enspace\alpha} U^J_{\enspace\beta}
    \,.
\end{equation}

\paragraph{Example (high energy limit of $\boldsymbol{W\phi\phi}$-scattering):}
A less trivial example is the high-energy limit of a massive vector boson ($W$) with two massless scalars ($\phi$). 
Like the previous example, there is one independent term. 
However, this time, the leading term vanishes
\begin{align}\label{eq:amp100}
        A(W,\phi,\phi) 
        &= A (\mbf{1}^{\frac{1}{2}+\frac{1}{2}}, {2}^{0}, 3^0)
        = \, \langle \mbf{1}_{\alpha} | 3 | \mbf{1}_{\dot\alpha} \rbrack 
        \nn\\
        &\overset{m \ll E}{\longrightarrow}  \left(
        \langle k_{1,I} | 3 | k_{1,J} \rangle 
        + m\ \langle q_{1,I} | 3 | k_{1,J} \rangle
        - m\ \langle k_{1,I} | 3 | q_{1,J} \rangle 
        - m^2\ \langle q_{1,I} | 3 | q_{1,J} \rangle 
        \right)  
        U^I_{\enspace\alpha} 
        U^J_{\enspace\dot{\alpha}}
        \nn\\
        &= m \left[
        \langle q_{1,I} | 3 | k_{1,J} \rangle - \langle k_{1,I} | 3 | q_{1,J} \rangle 
        \right] 
        U^I_{\enspace\alpha} 
        U^J_{\enspace\dot{\alpha}}
        + \mathcal{O}(m^2) 
        \,.
\end{align}
The above expression is unsatisfactory 
since we use $|k_{1,I}\rangle$ to label the massless state after taking the high energy limit. 
Therefore, we need to express the spinor $|q_{1,I}\rangle$ in terms of $|k_{1,I}\rangle$.

To do this, we use the following identity
\begin{equation}\label{eq:q2qk}
    \vSpinor_{A}| q_{I} \rangle = \vSpinor_{A}| q_{J} \rangle \, \varepsilon^{JK} \, \langle q_{K} | k_{I} \rangle = \vSpinor_{A}| q | k_I\rangle \,,
\end{equation}
where $q^{\mu}$ should be regarded as a \emph{reference momentum} for the massless momentum $k^{\mu}$ with the associated reference spinor $|q_I\rangle$.
The spinor $|q_I\rangle$ must satisfy  (\ref{eq:qkrelation}) but is otherwise unconstrained. 
This gauge freedom can be parameterized as
\begin{equation}\label{eq:qredundancy}
    | q'_{I} \rangle = | q_{J} \rangle + | k_{J} \rangle {L^J}_I \,,
\end{equation}
where $L$ is an arbitrary little group transformation.

Applying (\ref{eq:q2qk}) to \eqref{eq:amp100}, one finds that the leading term in the high energy limit is a true massless amplitude that is gauge invariant
\begin{equation} \label{eq:Wphiphi_HE}
    A(W,\phi,\phi) \overset{m \ll E}{\longrightarrow} 
    m \left[ 
    \langle 1_{I} | q_1 | 3 | 1_{J} \rangle 
    - \langle 1_{I} | 3 | q_1 | 1_{J} \rangle 
    \right] + \mathcal{O}(m^2)
    \,,
\end{equation}
where we have dropped the $U$-matrices. 
Gauge invariance is guaranteed because $(q_1)_A^{\enspace B}$ is always contracted into a $\la 1_I \vert^A$.

As a sanity check, we provide an additional example illustrating how terms conspire to keep the leading term of the high energy limit gauge invariant.

\paragraph{Example (high energy limit of $\boldsymbol{WW\phi}$-scattering):}

There are two independent terms in this case
\begin{equation}
    \begin{aligned}
        A(W,W,\phi) 
        = A(\mbf{1}^{\frac{1}{2}+\bar{\frac{1}{2}}}, \mbf{2}^{\frac{1}{2}+\bar{\frac{1}{2}}}, 3^0) 
        = g_1 \, {}_{\alpha_1}\langle {\bf 12} \rangle_{\beta_1} \, {}_{\dot{\alpha}_2} \lbrack \mathbf{12} \rbrack_{\dot\beta_2} + g_2 \, {}_{\alpha_1}\langle \mathbf{12} \rbrack_{\dot\beta_2} \, {}_{\dot{\alpha}_2} \lbrack \mathbf{12} \rangle_{\beta_1} \,.
    \end{aligned}
\end{equation}
It is useful to make a change of basis from the couplings $(g_1,g_2)$ to $(g_+,g_-) = (g_1+g_2, g_1-g_2)$ and study the high energy limit of the $g_\pm$ components.

In the high energy limit, the $g_+$ component amplitude reduces to 
\begin{equation}
\label{eq:g+}
    A(W,W,\phi)|_{g_-=0} \overset{m \ll E}{\longrightarrow} 
    \langle 1_{I_1} 2_{J_1} \rangle 
        \langle 1_{I_2} 2_{J_2} \rangle 
    + \langle 1_{I_1} 2_{J_2}  \rangle
        \langle 1_{I_2} 2_{J_1} \rangle
    \sim u_{1,I_1} u_{1,I_2} u_{2,J_1} u_{2,J_2} \,,
\end{equation}
which is exactly the three-point massless amplitude with spin (1,1,0). 

On the other hand, the $g_-$ amplitude is much more complicated
\begin{align}
        A(W,W,\phi)|_{g_+=0}
        \overset{m \ll E}{\longrightarrow} \, & \Big(
            \langle 1_{I_1}| 2_{J_1} \rangle 
            {+} m \langle 1_{I_1} | q_1 | 2_{J_1}  \rangle 
            {+} m \langle 1_{I_1} | q_2 | 2_{J_1}  \rangle 
            {+} m^2 \langle q_{1,I_1} | q_{2,J_1} \rangle 
        \Big)
        \\ \times &
        \Big(
            \langle 1_{I_2} | 2_{J_2} \rangle 
            {-} m \langle 1_{I_2} | q_1 | 2_{J_2} \rangle 
            {-} m \langle 1_{I_2} | q_2 | 2_{J_2} \rangle 
            {+} m^2 \langle q_{1,I_2} | q_{2,J_2} \rangle 
        \Big) - (I_1 \leftrightarrow I_2)
    \nn 
\end{align}
While the leading term vanishes,
the subleading term is a gauge invariant
\begin{equation}
    \label{eq:g-}
    \begin{aligned}
    A(W,W,\phi)\vert_{g_+=0} 
    \overset{m \ll E}{\longrightarrow} 
        & - m \langle 1_{I_1} 2_{J_1} \rangle \Big(
            \langle 1_{I_2} | q_1 | 2_{J_2} \rangle + {}_{I_2}\langle 1 | q_2 | 2 \rangle_{J_2} 
        \Big) 
        + (I_1, J_1)  \leftrightarrow (I_2, J_2)
    \\
        &\quad = \, -2m \, \langle 1_{I_1} | q_1 | 2 | 1_{I_2} \rangle \, \varepsilon_{J_1,J_2} 
        + 2m \, \langle 2_{J_1} | q_2 | 3 | 2_{J_2} \rangle \, \varepsilon_{I_1,I_2} \,.
    \end{aligned}
\end{equation}

Note that the first and second terms are precisely the spin-(1,0,0) and spin-(0,1,0) three-point massless amplitudes, respectively. 
It is not surprising that the \textit{massive} five-dimensional spin-(1,1,0) amplitude contains both the \textit{massless} spin-(1,0,0) and spin-(0,1,0) amplitudes in its high energy limit because the little group for massive particles lives in four dimensions, while the little group of massless particle lives in three dimensions. 
Therefore, a four-dimensional vector should reduce to a vector$\,\oplus\,$scalar in three dimensions. 

\paragraph{The mass dimension and spin drop.}
As exemplified by \eqref{eq:Wphiphi_HE} and \eqref{eq:g-}, the mass dimension of a massive amplitude is not preserved by the high energy limit. 
In addition, the spin of a massive amplitude can also drop after taking the high energy limit. 
To see how mass dimension and spin drops arise, we consider the high energy limit of antisymmetrization and symmetrization of the little group indices of $\langle \mathbf{1}_\alpha | \lbrack \mathbf{1}_{\dot{\alpha}} |$.

We start by computing the high energy limit of antisymmetrization 
\begin{align}\label{eq:antisymlimit}
        & \langle \mathbf{1}_{\alpha} |_A \, \lbrack \mathbf{1}_{\dot\alpha} |_{B} - \lbrack \mathbf{1}_{\dot\alpha} |_{A} \, \langle \mathbf{1}_{\alpha} |_{B} 
        \nn\\ & \qquad
        \overset{m \ll E}{\longrightarrow}  \Big( \langle 1_{I} |_{A} + \langle 1_{I} | q |_{A} \Big) \Big( \langle 1_{J} |_{B} - \langle 1_{J} | q |_{B} \Big) - \Big( \langle 1_{J} |_{A} - \langle 1_{J} | q |_{A} \Big) \Big( \langle 1_{I} |_{B} + \langle 1_{I} | q |_{B} \Big) 
        \nn\\ & \qquad
        = \varepsilon_{IJ} (p_1 )_{AB} + m (\varepsilon_1^{i})_{AB} \cdot \Gamma_{IJ}^i + m^2 \, \varepsilon_{IJ} (q)_{AB} \,,
\end{align}
where the $\Gamma^i_{IJ}$ are $\SO(3)$ $\Gamma$-matrices. 
The first term above the leading term that corresponds to the scalar component in the high energy limit. 
The second term is subleading and corresponds to the vector component in the high energy limit. 
The third subsubleading term is not gauge invariant and cannot appear as the leading high energy limit in a physical amplitude.

On the other hand, the symmetrization is 
\begin{align}
        \langle \mathbf{1}_{\alpha} |_A \, \lbrack \mathbf{1}_{\dot\alpha} |_{B} {+} \lbrack \mathbf{1}_{\dot\alpha} |_{A} \, \langle \mathbf{1}_{\alpha} |_{B} 
        &\overset{m \ll E}{\longrightarrow}  \Big( \langle 1_{I} |_{A} \, \langle 1_{J} |_{B} + \langle 1_{J} |_{A} \, \langle 1_{I} |_{B} \Big) 
        + m \Big( {}_{A}| 1 | q |_{B} {-} {}_{A}| q | 1 |_{B} \Big) \varepsilon_{IJ} 
        \nn\\&\qquad
        + m^2 \Big( \langle q_{I} |_{A} \, \langle q_{J} |_{B} + \langle q_{J} |_{A} \, \langle q_{I} |_{B} \Big)
\end{align}
where ${}_{A}| 1 | q |_{B} = \vSpinor_{A}| 1_I \ra \la 1^I | q_J \ra {\la q^J |}_{B}$. 
Here, the first term corresponds to the vector component and is the only gauge-invariant term. 
All other terms are subleading and not gauge invariant.

There are two important lessons from the above derivation. 
First, because explicit factors of $m_i$ can absorb mass dimensions, the high energy limit of massive amplitudes can contain components with lower mass dimensions.
Second, the high energy limit of massive amplitudes also contains components with lower spin. 
We can also see that in massive amplitudes with antisymmetrization (whether manifest or not\footnote{The non-manifest antisymmetrization means the rest part of the amplitude is antisymmetric in Lorentz indices, which will only keep the antisymmetric part of the spinors $|\boldsymbol{i}\rangle |\boldsymbol{i}\rbrack$.}) on chirality change, it's not straightforward to analyze its mass dimension under high energy limit. It also depends on the rest of the amplitude: what the antisymmetrized spinors contract with and whether the leading term in (\ref{eq:antisymlimit}) vanishes in the final amplitudes.

While the mass dimension of massive amplitudes is not well defined, the mass dimension of their high energy limit is. 
Therefore, one can categorize the possible independent terms in amplitudes by the number of symmetrizations/antisymmetrizations performed when taking the high energy limit. 
Roughly speaking, each antisymmetrization lowers either the mass dimension or spin by one, while symmetrization preserves both mass dimension and spin (in some cases, it can do both).

\acknowledgments 

We are grateful to 
Nima Arkani-Hamed, Shounak De, Taylor Knapp, Marcus Spradlin, and Akshay Yelleshpur Srikant for useful discussions.  This work was supported in part by the US Department of Energy under contract DE-SC0010010 Task F (AP, LR, AV), by Simons Investigator Award \#376208 (AV), and by Bershadsky Distinguished Visiting Fellowship at Harvard (AV). S. Rajan was also supported by a Karen T. Romer Undergraduate Teaching and Research Award.

\appendix

\bibliographystyle{JHEP}
\bibliography{reference}

\end{document}